\documentclass[graybox]{svmult}

\usepackage{type1cm}        
\usepackage{makeidx}         
\usepackage{graphicx}        
\usepackage{multicol}        
\usepackage[bottom]{footmisc}

\usepackage{newtxtext}       %
\usepackage[varvw]{newtxmath}       

\makeindex             

\begin{document}

\title*{Blockchain in a nutshell}
\author{Duc A. Tran and Bhaskar Krishnamachari}
\institute{Duc A. Tran (corresponding author) \at University of Massachusetts, 100 Morrissey Blvd, Boston, MA 02125, \email{duc.tran@umb.edu}
\and Bhaskar Krishnamachari \at University of Southern California, 3740 McClintock Avenue, Los Angeles, CA 90089, \email{bkrishna@usc.edu}}
%
%
\maketitle

\abstract*{Blockchain enables a digital society where people can contribute, collaborate, and transact
without having to second-guess trust and transparency. It is the technology behind the
success of Bitcoin, Ethereum, and many disruptive applications and platforms that have
positive impact in numerous sectors, including finance, education, health care,
environment, transportation, and philanthropy, to name a few. This chapter provides a friendly description of essential concepts,  mathematics, and methods that lay the foundation for blockchain technology.
}

\abstract{Blockchain enables a digital society where people can contribute, collaborate, and transact
without having to second-guess trust and transparency. It is the technology behind the
success of Bitcoin, Ethereum, and many disruptive applications and platforms that have
positive impact in numerous sectors, including finance, education, health care,
environment, transportation, and philanthropy, to name a few. This chapter provides a friendly description of essential concepts,  mathematics, and algorithms that lay the foundation for blockchain technology. }

\section{Introduction\label{sec:1}}
Let us consider the following favorite game of our childhood: Alice and Bob each bet \$100 on the outcome of a coin toss, whether it is ``head'' or ``tail''. Alice calls the outcome and Bob is the tosser. Alice will win the bet if her guess is correct and Bob will otherwise. It is so easy and simple a game, isn't it? Not really. What if Alice and Bob play this game remotely or separated by a brick wall such that Alice does not see the toss? How can Alice trust that Bob is honest? Bob can easily cheat;  knowing Alice's prediction he can say the opposite outcome. Even in the case he is honest, Alice may not be. She can run away not giving Bob the \$100 she bet, assuming she sprints so fast that he cannot catch her.

The above game is an example of a big real-world problem  we see almost everywhere. That is, how to quickly process transactions for everybody, possibly involving multiple people, in an environment not always honest, where people may not trust one another?

Our society has hundreds, thousands, of years been relying on the intermediaries to solve that problem. If we do not trust each other, let us do the transaction through a trusted middleman, hence the existence of banks for financial activities, central servers for  storage and computation, or, at a larger scale, central governments for maintaining the society. The trust put on the intermediaries is an \emph{assumed} trust: we assume that they will do what they are supposed to do. That is the perfect scenario, which is not the case in practice. Mistakes are made by humans. Machines  fail. Hackers are always looking for ways to penetrate into systems. Even in an ideal world where such errors or attacks do not happen, the conventional way of relying on a central authority to store information, process transactions, or manage activities for many people and institutions cannot scale. The   authority is  the bottleneck. It is increasingly expensive in both money and time  when there are more workloads.

This is where blockchain comes in. It is completely decentralized  with no intermediary involved. Blockchain overcomes the weaknesses of the centralized intermediary approach in four crucial aspects: trust, security, privacy, and transparency. Blockchain is trust-less; there is no need to raise the trust question. Bob and Alice in the aforementioned betting game do not have to worry about the other cheating.  Blockchain is  secure; while a central server as a single point of contact can be attacked or the data therein stored may maliciously be altered, blockchain as a system always functions correctly 24/7. As identity privacy is of   utmost importance today, it can be leaked in a middleman-based system. Blockchain does not allow this to happen as it is designed to hide personal identities. Lastly, about transparency, while today's banks may not disclose to us what they do behind  closed doors with our deposited money, blockchain makes all the transactions visible and verifiable. Since there is no concept of personal identity on the blockchain, making transactions visible does not cause loss of privacy.

Blockchain is capable to provide the above desirable properties thanks to its architecture as a decentralized network utilizing many computers owned by people. These computers collectively  store and process transactions in a way that although working autonomously they can still achieve consensus in decision making and be robust against malfunctions, attacks, dishonesty, and self-interests.
On the surface, we can think of blockchain as an Internet-like infrastructure for processing transactions. Using the Internet, one can send data from one computer to another without having to worry about how the data finds its way to get delivered or  whether the data can be lost; the Internet takes care of all those things so that we can focus on the main business job. Similarly, if people transact on the blockchain, they do not have to worry about many what-ifs, including trust about whether the other side may act as agreed upon or whether  money may be lost or  data maliciously changed. Blockchain has its name because, as a digital ledger, the transactions are stored in blocks, each new block appended to the previous to form a chain; hence the name \emph{blockchain}. Two consecutive blocks are mathematically linked in such a way that any change in an existing block would violate the mathematics of the  link with the next block. The mathematical methods used for this linking are from the field of mathematical cryptography, hence the name \emph{crypto} in ``cryptocurrencies" we see trending today. 

Trust is the biggest bottleneck in realizing transactions. It is the biggest bottleneck in advancing the society. As a trust-less system, blockchain removes that bottleneck. It makes sense that many consider blockchain the next big thing since the birth of the Internet. The Internet removes the geographical constraint, moving people closer for communication despite geographical distances. Blockchain, by removing the trust distance, moves people closer for doing actual transactions. Putting blockchain together with AI, a field of great mention today, we can think of AI as the brain of a system whereas Blockchain is the body. AI needs computing resources and training data to realize its promise. Blockchain is no less important because it is the best way to incentivize people to contribute computing power and good data,  the only way if we care about trust, security, privacy, and transparency. 

Blockchain is still in an early application stage. The space for blockchain-based developments is immense. To consider whether blockchain may apply to your business, at least four out of six following conditions should hold: 1) data is shared by multiple parties, 2) data is updated by multiple parties, 3) verification is required, 4) it is expensive to rely on intermediaries, 5) valid transactions must be eventually executed, and 6) transactions are inter-related. Most applications satisfy this, which are found  in almost every sector, including financial services, product manufacturing, energy and utilities, healthcare, e-government, retail and consumer, entertainment and media, just to name a few. 

According to Harvard Business Review \cite{IansitiHBR2017}, one can argue that Blockchain is not only a disruptive technology, but  has the potential to create new foundations for our economic and social systems; it is a foundational technology.  A recent PwC report \cite{pwc2020timefortrust} projected that Blockchain by 2030  will potentially add 1.76 trillion USD to the global GDP,  create 40 million new jobs, and be used to support 10\%-20\% of global business infrastructures. The 2020 annual global blockchain survey of Deloitte \cite{deloitte2020blockchain} interviewing 1488 business leaders from 14 countries, who had certain knowledge about Blockchain, found that 
39\% of the businesses applied Blockchain, a 23\% increase from 2019, 
55\% considered Blockchain a top-5 priority, and 
82\% would hire blockchain staff within 12 months.

\section{What is Blockchain}
\label{sec:2}
\begin{figure}[t]
\centering
\includegraphics[width=\textwidth]{ 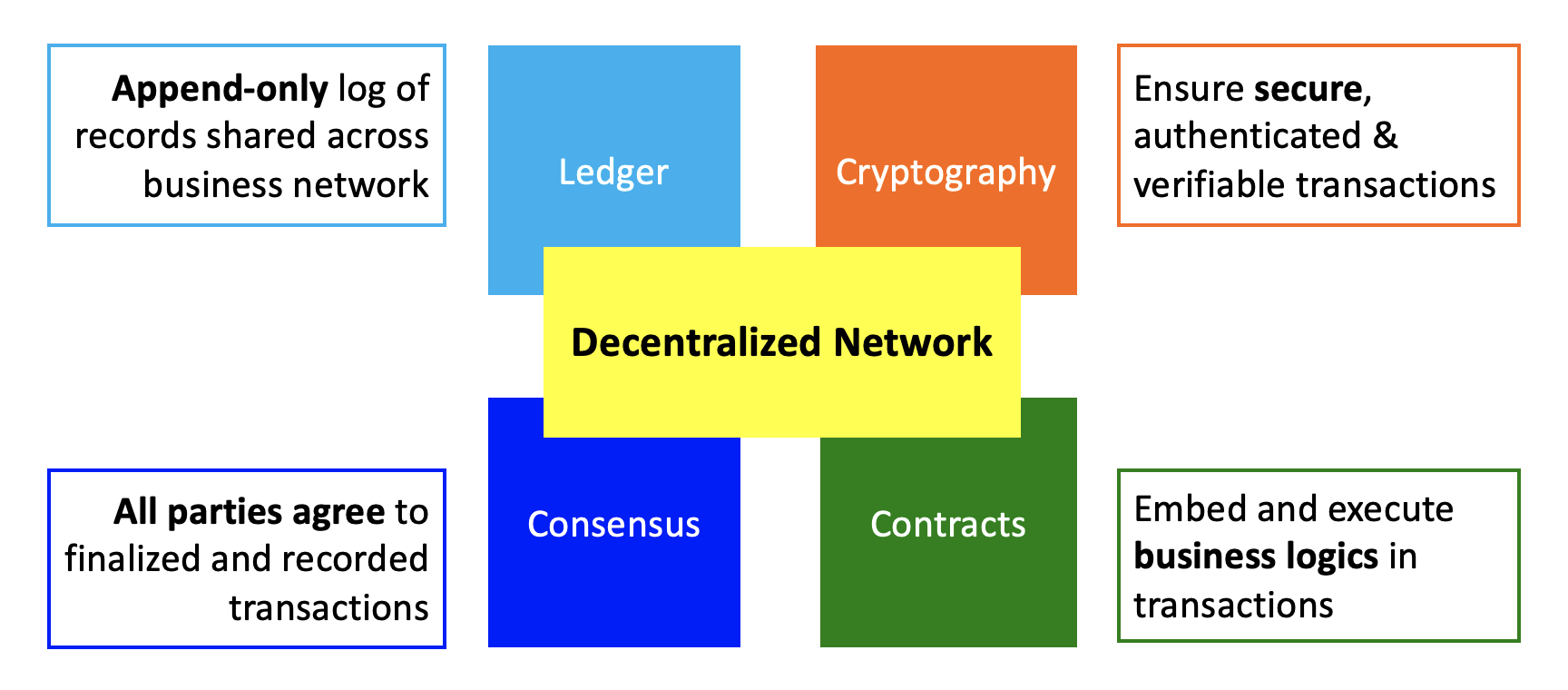}
\caption{The five constituent components of blockchain: decentralized network, cryptography, consensus, ledger, and contracts.}
\label{fig:components}
\end{figure}

Having introduced the motivation for Blockchain and its potentials, we now focus on what it actually is. To non-technical people, one can define Blockchain based on what it offers: a computing technology for transaction recording and processing that is
safe (no loss or mutability of data possible), transparent (easy verification and tracing), and trust-less (confidence of transacting without any intermediary). Technically, the most complete definition of Blockchain should see it as a decentralized computing system 
of five constituent components: decentralized networking, mathematical cryptography, distributed consensus, transaction ledger, and smart contracts, as illustrated in Figure \ref{fig:components}:
\begin{itemize}
\item \underline{Decentralized networking}: For computing, Blockchain relies on a decentralized network of computers, called blockchain nodes, that contribute computing resources to help store and process transactions. These computers work autonomously and communicate with each other in a peer-to-peer (P2P) manner. Most blockchain networks including Bitcoin adopt an unstructured P2P topology; i.e., a node chooses its neighbors arbitrarily. Some  networks such as Ethereum  use a structured one like Kademlia Distributed Hash Table \cite{maymounkov02} to optimize the P2P communication. Unstructured P2P may be less efficient than structured P2P, but the latter is more difficult to maintain, especially in a permissionless blockchain. Ethereum uses Kademlia but only as an add-on assistance \cite{DBLP:journals/tnse/WangZYZL21}; in other words, it still works with any unstructured P2P topology, albeit less efficient if only so.
\item \underline{Mathematical cryptography}: Cryptographic methods used in blockchain provide mathematical proofs that the blockchain must function as supposed to. Cryptographic hash is used to link data blocks in the  chain so that no data alteration is allowed post recording into the blockchain.  Each transaction is encrypted with Public-Key cryptography to ensure that the sender is verifiable using digital signature and only the intended recipient of the transaction can be the receiver.   Transaction confidentiality is achieved thanks to the method of Zero Knowledge Proof \cite{DBLP:conf/stoc/BlumFM88}. The choice of cryptography  to use determines the performance and guarantees of the blockchain. For example, Dogecoin blockchain clones  Bitcoin but using simpler cryptographic functions to increase transaction throughput; the mining in Dogecoin is based on SCRYPT which is faster and easier to run than SHA256 used in Bitcoin. This, however, results in weaker security, less robust to attacks by dishonest nodes.
\item \underline{Transaction ledger}:  As a storage technology, Blockchain is a digital ledger that stores the transactions  chronologically in blocks which are added in an append-only manner. This is the default data structure of the ledger for almost all blockchain networks. However, some blockchain networks, for example, Hedera \cite{hedera} and Fantom \cite{https://doi.org/10.48550/arxiv.2108.01900}, design the  ledger as a directed-acyclic graph (DAG) of blocks (or transactions) instead of a chain structure which can only append blocks. A chain is  a simple case of DAG because it shares the property of being directed-acyclic. The former offers simplicity but the latter is more efficient in transaction processing (for example, searching for a transaction is faster). The ledger structure, the block structure, and the number of transactions  in a block  are important  considerations when designing the ledger component of the blockchain. 
\item \underline{Distributed consensus}:  When a decision needs to be made, for example whether a transaction is valid, there is no central authority to decide. Instead, the decision is made based on consensus reached among the participating nodes.  Therefore, a blockchain network must have a consensus protocol to make sure that every  transaction or block  added to the blockchain is the one and only version of the truth that is agreed upon by all the nodes. Proof-of-Work consensus \cite{bitcoin}, giving more decision power to  nodes with more hardware-computing power, is adopted in early blockchain networks (Bitcoin, Litecoin, Ethereum in its original version).  Proof-of-Stake consensus \cite{10.1145/3132747.3132757}, giving more decision power to nodes with more financial stake, is popular among today's blockchain networks; its first functioning use for cryptocurrency was in Peercoin in 2012 \cite{peercoin}. The choice of  consensus protocol  is the most critical consideration in designing a blockchain network.
\item \underline{Smart contracts}: A blockchain can be considered a non-conventional kind of computers to perform certain tasks. Instead of being a computer integrating built-in computing processing units (the CPUs), blockchain is a decentralized computer utilizing hundreds or thousands of computers anywhere in the world. Applications that run on the blockchain are implemented as ``smart contracts", a term coined by Nick Szabo in the 1990s \cite{smartcontract}. A smart contract is nothing but a computer program; the term is used  because an application deployed on the blockchain  always functions correctly as programmed, like executing the conditions in a legal contract. This contract is smart because of its automated execution without human intervention. 
\end{itemize}
Next, we elaborate further on these components and their importance. We do not attempt to cover every aspect and every detail. Instead, we select certain issues to discuss hoping that the reader can have a quick understanding of what blockchain is and requires. More details will follow later to dig deeper into the technicality of blockchain.

\subsection{The Blockchain Computer}
\begin{figure}[t]
\centering
\includegraphics[width=0.9\textwidth]{ 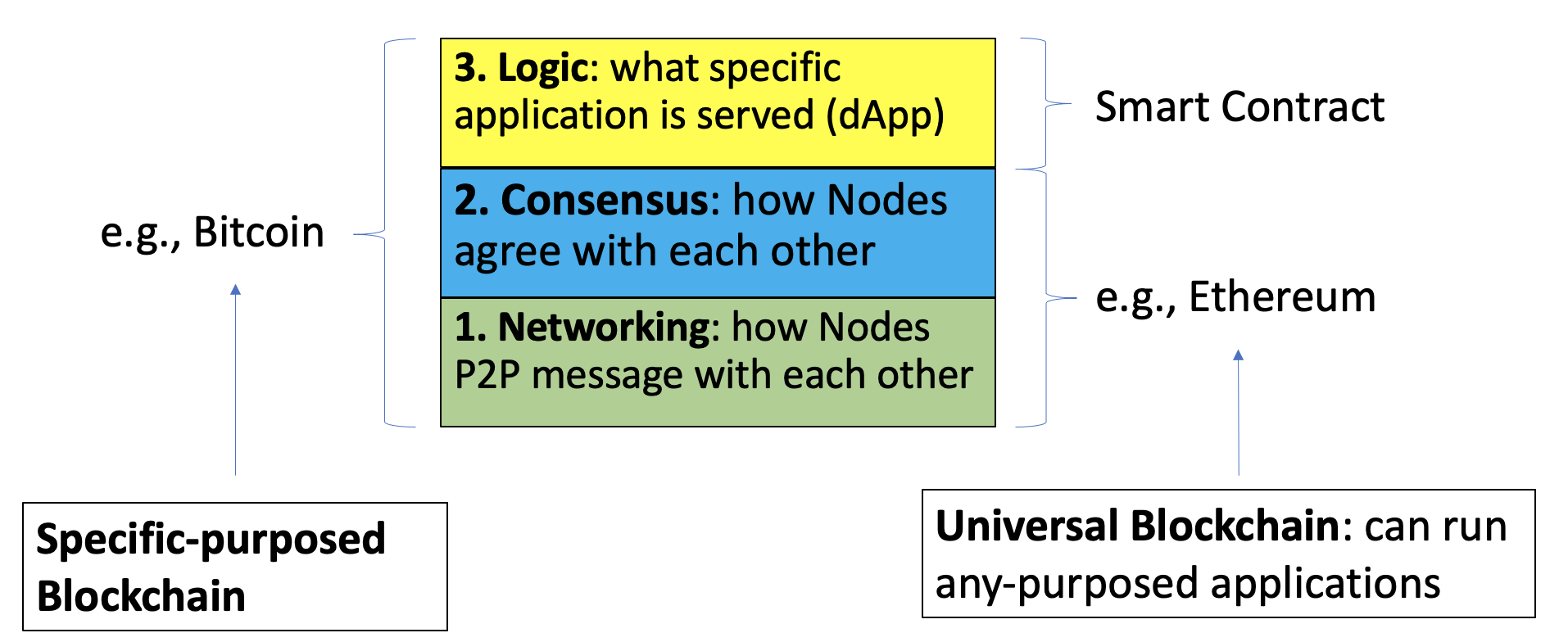}
\caption{Architecture of Blockchain as a new kind of computer.}
\label{fig:architecture}
\end{figure}

We can view Blockchain as a computer whose  architecture consists of three layers, illustrated in Figure \ref{fig:architecture}: the P2P networking layer, the consensus layer, and the logic layer. For example, Bitcoin is a blockchain computer that implements all these layers, whereas Ethereum implements  the first two layers, leaving the logic layer to application developers. 
Bitcoin is a purpose-specific blockchain computer that performs only one application:  create a digital currency, the Bitcoin cryptocurrency as we all know, and functions for moving this currency between accounts. This application is a built-in logic of the Bitcoin blockchain, and as such smart contract is not a concept of Bitcoin. On the other hand, Ethererum is a universal blockchain computer; it was designed to enable deployment of arbitrarily-purposed applications on the blockchain. Therefore, Ethereum is called a smart-contract blockchain network. In contrast, Bitcoin is an application-specific blockchain, precisely a cryptocurrency blockchain.

Viewing Blockchain as a computer is an intuitive observation. Essentially, a computer  is a machine that automates processing of applications, and it is thus reasonable that blockchain can be seen as a computer, at least virtually. In early years, with Desktop Computing, we have applications running on a desktop computer near us, in our home or office; we control this desktop computer. The past decade has seen many businesses moving to Cloud Computing; the cloud provider controls the ``cloud computer" (AWS cloud of Amazon or Azure Cloud of Microsoft). The future, very soon, we argue  will be the era of Blockchain Computing; nobody controls the blockchain computer.

This is a natural evolution in computing. Cloud Computing has replaced Desktop Computing to reduce the cost to maintain the IT system for  businesses and at the same time more efficiently utilize   computing resources. It is a one-stop shop  to satisfy all computing needs so that  companies can focus more time on their business logic. Compared to Cloud Computing,  Blockchain Computing offers the benefit of decentralization and trust guarantees. The cloud provider  has the power to manipulate the cloud computer; we have to trust this organization. Blockchain Computing is trustless and anyone can be a part owner of it. 

\subsection{The Blockchain State}
To interact with the blockchain, one needs an address, or, interchangeably, an account. The blockchain state consists of the set of addresses and information about them. As the state changes from time to time, Blockchain can be modeled as a state machine. It starts with a genesis state (when the blockchain is launched) and transitions from one state to the next upon triggering events (when transactions are added to the blockchain). We need to keep track of the blockchain state at any point of time. Depending on how the blockchain is designed, the state's data structure may differ.
It can be transaction-based (the state information consists of the list of transactions) or account-based (the state information consists of account balances). The data structure to represent transactions can also vary. We compare these models below, assuming for simplicity that each transaction is a transfer of value (asset) between addresses.

\subsubsection{Transaction-based Model}

 In the transaction-based model,  known as  Unspent Transaction Output (UTXO) \cite{bitcoin} conceived by Bitcoin, each transaction can send value to one or more recipients. It   consists of the following information:
\begin{itemize}
\item \underline{Output field}:  A list of receiving addresses and the amount of fund to be sent to each respectively. Each transfer output is called a UTXO transaction.
\item \underline{Input field}: A list of   UTXO transactions that will provide the fund for the transaction. These UTXO's previously sent funds to the sender and currently are unspent.
\end{itemize}

\begin{figure}[t]
\centering
\includegraphics[width=0.83\textwidth]{ 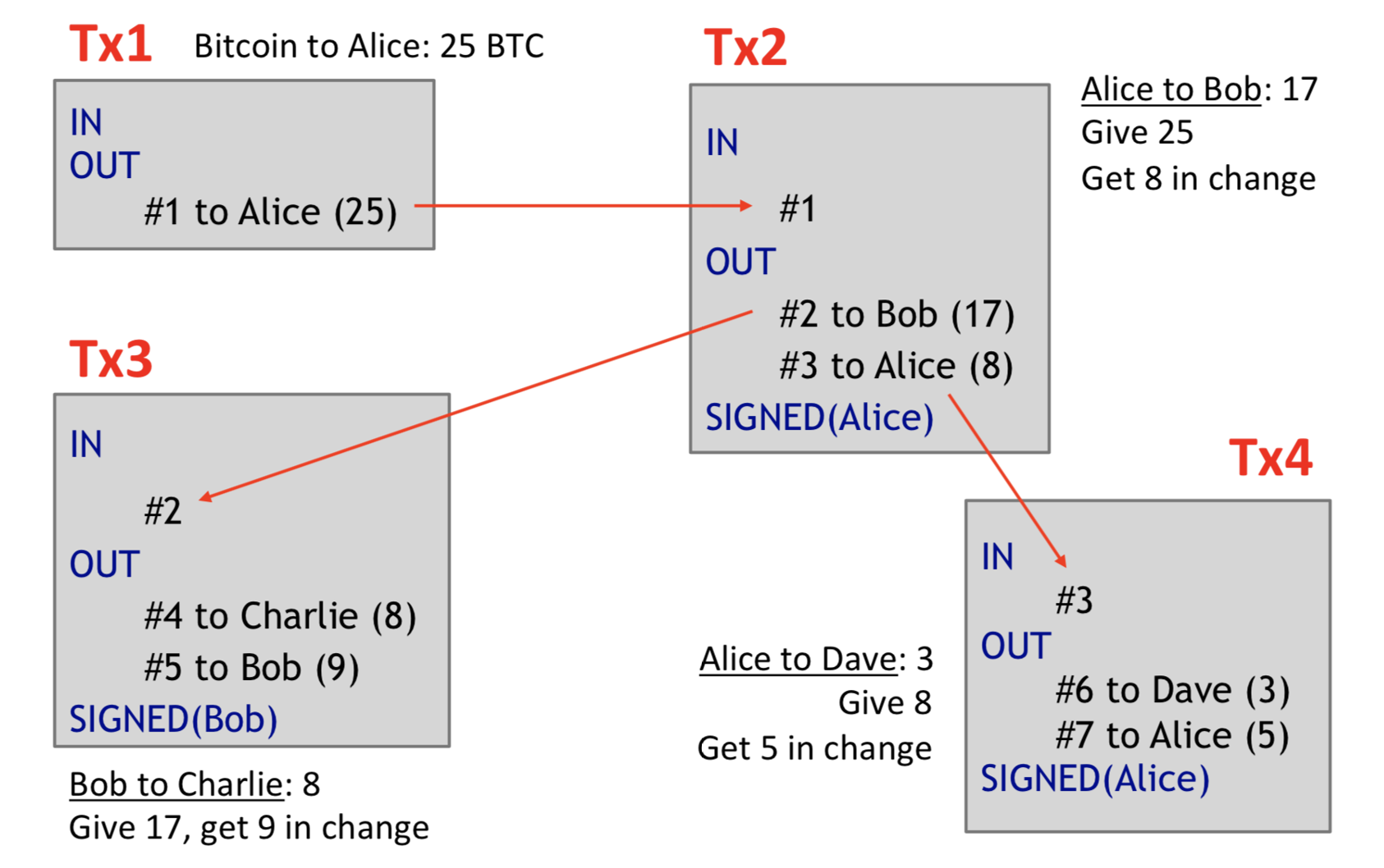}
\caption{The Unspent Transaction Output (UTXO) model: the blockchain state at the current time is the list of all unspent transactions. }
\label{fig:utxo}
\end{figure}

Figure \ref{fig:utxo} provides an example of Bitcoin transactions. The very first transaction \texttt{Tx1}, called the genesis transaction, sends 25 BTC to Alice. The input field is empty because this is the very first transaction of the blockchain operation, meaning Alice is the first recipient of Bitcoin (somebody has to be the first recipient). This transaction results in creation of a UTXO transaction, \texttt{Tx1}(\#1).  The second transaction \texttt{Tx2} is initiated by Alice, sending 17 BTC to Bob and the rest, 8 BTC, to herself. The total fund to send, $17+8=25$ BTC, comes from the fund that Alice previously received in UTXO \texttt{Tx1} (\#1). Because \texttt{Tx1} (\#1) is unspent, she has enough money for \texttt{Tx2}. After this execution, UTXO \texttt{Tx1} (\#1) is marked as ``spent'' and new UTXO \texttt{Tx2} (\#1, \#2), are created and marked as ``unspent''. Later, Bob initiates transaction \texttt{Tx3} to send 8 BTC to Charlie, with the remaining 9 BTC to himself. The total fund to send, $8+9=17$ BTC, comes from the fund that he previously received in UTXO \texttt{Tx2} (\#2). Because \texttt{Tx2} (\#2) is unspent, he has enough money to execute \texttt{Tx3}. After this execution, UTXO \texttt{Tx2} (\#2) is marked as ``spent'' and new UTXO \texttt{Tx3} (\#4, \#5) are created.

The blockchain state is the set of current UTXO transactions. Each time a UTXO transaction is used as an input in a new transaction, the input UTXO will be marked as ``spent"  thus no longer usable and each output  sending fund out  will be created as a new UTXO transaction. The new UTXO transaction(s) may be used later as input providing funds to future transactions. The marking of input UTXO transactions as ``spent" is to avoid double spending, which means spending the same UTXO for two different transactions.  The UTXO blockchain state does not directly provide account balances. To know how much Alice has in her account, one needs to sum all the funds she received in current UTXO transactions. 

In the case that the total input fund has more  than the output, the remaining balance can be sent to the sender's own address to avoid losing fund. For example, in transaction \texttt{Tx4}, Alice sends 3 BTC to Dave out of the 8 BTC she has available from UTXO \texttt{Tx2} (\#3), but because UTXO \texttt{Tx2} (\#3) will be marked as ``spent'', in order not to  lose the $8-3=5$ BTC she has remaining, she creates a new UTXO  \texttt{Tx4} (\#7) to send this 5 BTC to herself. She does not lose any money. In some blockchains, for example, Bitcoin and Ethereum, Alice may not send all of the remaining balance to her address; in this case, the leftover will be sent as reward to the blockchain node that adds this transaction to the blockchain.

\subsubsection{Account-based Model}
The account-based model \cite{ethereum} is more intuitive. It is like the account model of a bank.  The state consists of the balance information for each address. When there is a transaction, the  balances of the sender's and receiver's accounts will be updated immediately and saved in the state. Therefore, when queried the account balance of an address is instantly available without any computation. 

A transaction in the account-based model is much simpler than a UTXO transaction. The former  consists of only one receiving address  and the  amount of fund to send. It is much faster to verify if the sender has enough fund, which is done by simply comparing two numbers: whether the sender's balance exceeds the  amount to send. In contrast, UTXO requires searching the blockchain state to see if the input UTXO's are indeed unspent. Consequently,  the account-based model offers a clear advantage when it comes to enabling  ``smart contracts''  (computer programs to deploy applications on the blockchain). For smart contracts, a transaction can be not only a transfer of value, but also a  call to a function of  arbitrary logic; it contains code data for executing this function. To process a transaction  thus involves   execution of the code in the transaction. As smart contracts are computationally expensive, simplicity of computation is important.  UTXO creates computational overhead because all spending transactions must be explicitly recorded. 

UTXO is suitable for a cryptocurrency blockchain like Bitcoin which serves only one application: transfer of money. Computation is not that complex. Another reason is due to transparency and traceability. Back to Figure \ref{fig:utxo}, if we want to know how Dave received 3 BTC from Alice in transaction \texttt{Tx4}, we can trace all the way to the beginning how the fund started and flowed. We can find that it started from Alice in \texttt{Tx1} (\#1) to  Alice in \texttt{Tx2} (\#3) to Dave in \texttt{Tx4} (\#6). In other words, every transfer has a non-fungible path. With the account-based model, if Dave received 3 BTC from Alice, this fund is fungible; we only know that this 3 BTC came  from Alice, not knowing any further where this particular 3 BTC arrived at Alice. In other words, UTXO is more transparent. That said, one could argue that the account-based model offers better privacy.

\subsection{The Chain Structure}
By default, and adopted in all but a few unpopular blockchain designs, the blockchain ledger  follows a chain structure. The data 
is organized into a chain of data blocks: $b_1$, $b_2$, $b_3$, ...When new transactions need to be saved, they are put in a new block which will be appended to the last block of the existing chain. In an account-based blockchain, e.g., Ethereum, a block also contains the blockchain state information (the balances of all the accounts at the current time).

Besides storing the transaction data, blockchain state if applicable, and necessary header information, the block has two important attributes:
\begin{itemize}
\item \underline{Block ID} $b_{i}.id$: This is set to the hash value of the  block content using a cryptographic hash function $H$; i.e., $b_{i}.id = H(b_i)$. This hash function is predefined and publicly known. 
\item \underline{Previous hash} $b_i.prev$:  This is set to the ID of the previous block $b_{i-1}$ to which $b_i$ is appended; i.e., $b_i.prev = b_{i-1}.id$. 
\end{itemize}
It is noted that the block ID may not necessarily be stored in the block because it   can be computed from the block's content.

The previous-hash information is critical in maintaining the data integrity of the chain. If any part of any block  is changed after it is recorded in the blockchain, this will be detected. This is because for a new block to be added to the blockchain it must pass a procedure called \emph{block validation}.  A new block $b_{i+1}$ is valid if and only if 
\begin{enumerate}
\item Previous-hash is consistent: $b_{i+1}.prev = H(b_i)$ 
\item All the transactions in $b_{i+1}$ are valid
\item Previous block $b_i$ is valid
\end{enumerate}

Let us put Step 2 aside (to be discussed later). The verification in Step 1 requires computing the hash value of $b_i$ and comparing it with $b_{i+1}.prev$. Step 3 requires  running the same block validation procedure to verify the validity of block $b_i$. Consequently, the validation procedure for block $b_{i+1}$ requires  checking whether the previous-hash value stored in  block $b_j$   equals the hash value of its previous block $b_{j-1}$ for all   $j \le i+1$. If an earlier block, say $b_{j-1}$, has been changed from its original value, when we compute its hash value, $H(b_{j-1}$), we will find it not identical to the previous-hash value $b_{j}.prev$ stored in block $b_j$.  This is a violation and as a result the new block $b_{i+1}$ is concluded to be  invalid and not added to the blockchain.

A consequence of block $b_j$ being changed is that the blockchain will never grow beyond the time of this change. One might say, ``that means, the blockchain is useless then, because just one block's modification halts the whole blockchain''. This is true if the blockchain network consists of only one computer. In practice, the blockchain network runs many computers, where the blockchain data is replicated on every computer node. For a node to ensure that its blockchain copy is correct (same as  the globally correct version), it needs to compare its copy with that  of the neighbors and choose to use the longest\footnote{Comparing based on blockchain length (the number of blocks in the blockchain) is adopted in most blockchain networks, but other comparison criteria have also been explored, for example, choosing the ``heaviest'' blockchain copy as the correct one, where  ``heaviness''  is a weighted-generalization of the length.} blockchain as the correct one. Before this comparison takes place, the node needs to check the validity of each neighbor's blockchain copy, which requires validating all the blocks in this copy. Therefore, if a blockchain copy from some node  contains a violation, this copy will fail the  validation step. As such, the bad copy will not be used by the honest nodes in the network.

\subsection{Use of   Cryptography}
It is now clear that the data immutability of the blockchain is achieved thanks to the previous-hash information linking between consecutive blocks in the blockchain. However, in theory, a hash function may have different input values resulting in the same hash output, meaning that block $b_{j-1}$ can be changed  from its original value such that its  hash value, $H(b_{i-1})$, remains the same as before, which  equals   the previous-hash value $b_j.prev$ stored in block $b_j$. In this case, the block validation procedure cannot detect the change. The choice of the hash function is therefore critical. We should choose one so that even though such a block alteration without being detected is theoretically possible, realizing it is practically impossible. For this reason, the hash function $H$ used in blockchain must be a \emph{cryptographic} hash function, not any arbitrary hash function.

Recall that a hash function is a one-way function that  takes an input of arbitrary length to output a string of constant length, here assuming that values are represented as binary strings. For example, SHA256 is a hash function that outputs a  binary string of 256 bits. A cryptographic hash function $H$ is a hash function with three properties: 
\begin{itemize}
\item \underline{Collision-resistant}: It is infeasible to find different input messages $x$ and $y$ such that $H(x) = H(y)$. 
\item \underline{Hiding}: Given the output $c = H(x)$, it is infeasible to find an input $x$.
\item \underline{Puzzle-friendly}: If we know the hash value $c = H( r \| x)$ of an input message made by concatenation of $r$ and $x$, and even if we know part of the input, $x$, we cannot reconstruct the remaining input $r$ in time complexity faster than $2^n$ where $n$ is the binary length of  output $c$.
\end{itemize}

Because of these properties, knowing $b_j.prev = H(b_{j-1})$, it is infeasible to find $b_{j-1}^{'}  \not \eq b_{j-1} $ such that $H(b_{j-1}^{'})$ = $H(b_{j-1})$. With $H$ being a cryptographic hash function, no one can alter an existing block not to be detected. The blockchain data is tamper-proof.

Cryptographic methods also have many other uses in the operation of a blockchain. Recall the coin bet  between Alice and Bob at the start of this chapter, in which a situation is what if Bob cheats. A cryptographic hash function $H$ can solve this cheating problem as follows:
\begin{enumerate}
\item \underline{Alice}: suppose that her prediction is $x$ (``head'' or ``tail'')
	\begin{itemize}
	\item Generate a secret random number $r$ (of some large binary length $n$).
	\item Compute $c = H(r \| x)$ (called ``prediction commitment").
	\item Send $c$ to Bob, instead of sending her prediction as raw data.
	\end{itemize}
\item \underline{Bob}: upon receipt of the prediction commitment $c$, he will send Alice the honest outcome $x^*$ of the coin toss. Because the hash function $H$ is cryptographic, he does not know the ground-truth prediction $x$ of Alice, and as such he has no reason to cheat.
\item \underline{Alice}: upon receipt of $x^*$, if her guess is correct, i.e., $x = x^*$ she will tell Bob that she wins by sending him the secret number $r$.
\item \underline{Bob}: upon receipt of number $r$, he will verify if the commitment $c$ he received earlier from Alice equals $H(r \| x^*)$ and convincingly accept the loss.
\end{enumerate}
This solution is called a Commitment Scheme in cryptography \cite{DBLP:conf/ac/Damgard98}. It is critical that the secret $r$ generated by Alice must come from a large number space. If  the binary length $n$ was small, it would take short time for Bob to exhaustively try all possible values of $r$ and combine with $x$=``head'' or $x$=``tail'' to see which combination satisfies $H(r \| x) = c$. When that combination is found, he can cheat by telling Alice that the outcome is the opposite value of $x$ found in this combination. When $n$ is large, even though $x$ can take only two possible values, ``head'' or ``tail'', Bob cannot reconstruct the secret $r$ thanks to the ``puzzle-friendly'' property of $H$ as a cryptographic hash function. 

The above is a glimpse into how mathematical cryptography helps make a system trustless. Alice and Bob do not need to question each other's honesty thanks to the Commitment Scheme. However, in the case Alice loses the bet,  what if she runs away? Intuitively, a solution is to at least require that they both have to deposit the bet money in a lockbox which when the outcome is announced will be unlocked to transfer all the money to the winner. This is to say that there is a lot more to do and mathematical cryptography is the main tool to realize all that.

\subsection{Where is Blockchain  Stored\label{sec: wheretostoreblockchain}}
As we explained earlier, the blockchain is a decentralized network of computers contributing computing resources to help with transaction storage and processing. Among these computers, where is the blockchain data  stored? Should we distribute the blocks in the blockchain ledger across these nodes so that some blocks are on node 1, some blocks on node 2, etc.? We should not because if node 1 fails, we cannot access the blocks stored there. Hence some redundancy is needed to guarantee availability; that is, a block should be replicated on more than one node. The next question then is, `` how much replication is enough?''. In blockchain, the blockchain ledger is  replicated fully on every node: each node stores a full copy of the entire blockchain. This is because of the blockchain's vision  to provide complete decentralization (no node depending on other nodes to access certain blocks) and complete availability (it is always accessible even in the worst case of failure). 

When a new node joins the blockchain network,  it must discover  existing nodes as neighbors and connect P2P  to them.  The new node obtains a blockchain copy from  these neighbors.  The list of blockchain nodes is available publicly. In most blockchain networks,  the P2P networking topology can be arbitrary; any existing  nodes can be selected at random, not geographically dependent. 

Over the time, since nodes work autonomously and independently, their local blockchain copies may disagree. To ensure consistency, they need to frequently, or upon some triggering event such as adding new transactions, send their blockchain copy to the neighbors or pull blockchain copies from the neighbors. When presented with multiple blockchain copies, a node must decide which copy is the globally correct one and uses it. As aforementioned, the default criterion is to choose the longest copy.

\subsection{ How to Process a Transaction\label{sec:processatransaction}}
When someone initiates a transaction  with the blockchain, this is usually done in a user-friendly front-end application that can interact with the blockchain network via API calls. This transaction needs to be sent to a blockchain node (in practice, multiple nodes in case one node may fail or behave wrongly) and will be processed as follows:
\begin{itemize}
\item Each node $X$ on first receipt of  transaction \texttt{Tx}:
	\begin{itemize}
	\item \underline{Transaction forwarding}: forward transaction \texttt{Tx} to the neighbor nodes of $X$.
	\item \underline{Transaction verification}: verify that the sender address of transaction \texttt{Tx} has sufficient fund to send. If so \texttt{Tx} is put into a mempool which is a queue of valid transactions waiting to be put in a new block.
	\item \underline{Blockchain creation}: pull pending transactions from the mempool to include in a new block $b$ and append this block to the existing blockchain ledger at  node $X$. Note that  block $b$ must include the previous-hash information (the hash value of the last block). 
	\item \underline{Block update}: send the new block $b$ to the neighbor nodes of $X$.
	\end{itemize}
\item  Each node $Y$ on first receipt of block $b$
	\begin{itemize}
	\item \underline{Block forwarding}: forward block $b$  to the neighbor nodes of $Y$.
	\item \underline{Block validation}: verify the validity of block $b$ on the existing blockchain ledger of node $Y$. This validation requires checking on the consistency of previous-hash information  and the validity of every  transaction in block $b$. 
	\item \underline{Block insertion}: append block $b$ to the blockchain ledger if it is valid. Else, ignore $b$.
	\end{itemize}
\end{itemize}

To validate a transaction during the Block Validation step above may vary from one blockchain design to another. In Bitcoin,  we only need to verify that the sender of the transaction has  available fund to spend. This verification is successful if the input  transactions exist in the blockchain state, meaning they are currently unspent, and the sum of output amounts in these  transactions is sufficient. However, in a smart-contract blockchain network like Ethereum, transaction validation may involve more work than just checking the balance sufficiency. If a transaction involves a function call to interact with a smart contract, the verification will need to run this function with the blockchain in the previous blockchain state (recorded in the previous block) and if the resulted blockchain state does not match the blockchain state recorded in the block under validation, the block is considered invalid.

The transaction processing procedure in blockchain is simple and allows for autonomous processing at the blockchain nodes. This simplicity, however, leads to several consistency problems. First, each transaction is broadcast to all the nodes and so the same transaction may be added to different blocks created at different nodes. We need to ensure that each  transaction can only be added to the blockchain once. 
Second, different nodes in parallel create different new blocks to attempt to append to the (same) existing blockchain. We need to ensure that only one of them will be added as the next block. Third, different nodes may have different copies of the blockchain. We need to ensure that they have to agree on a copy as the globally correct version. To resolve these inconsistencies,  the nodes have to regularly agree on the current state of the blockchain, and that is what we call consensus achievement. We need a consensus protocol.

\subsection{How to achieve consensus}

Consensus is a research area of computing with more than 30 years of study before blockchain became popular. It started in the 1970s with the NASA sponsored project, ``Software Implemented Fault Tolerance (SIFT)'' \cite{10.1145/1479992.1480025}, aimed to build a resilient aircraft control system. The challenge was to replicate the system on multiple machines such that the whole system can sustain multi-machine failures. The nominal work by Lamport et al. in 1982 \cite{10.1145/357172.357176} formulated this challenge as the now well-known ``Byzantine Generals' Problem" (BGP). It coined the notion of ``Byzantine Fault" to model a condition in a distributed system where  some  nodes are unreliable and  may appear arbitrarily normal or malicious and collude with each other such that there is no consistent information for the other nodes to declare their malfunction. 

A  Byzantine Fault Tolerance (BFT) system must avoid complete failure and for that the nodes must agree on a concerted strategy and live by this consensus, knowing that some nodes may fail or act maliciously. 	
BGP laid the foundation for research in distributed consensus. Companies like Google and Facebook started adopting scientific results in BFT consensus  for   mission-critical services such as Google Wallet and Facebook Credit.  The birth of Bitcoin in 2009 \cite{bitcoin} was the first time that consensus is realized in a large-scale practical environment in a permissionless and decentralized manner. The distributed consensus implementation by NASA, Google, or Facebook is not fully decentralized nor permissionless because the participating computers are controlled by these organizations. The Bitcoin network is public, requiring no permission for computers to participate and no centralized authority to make decisions.

To describe BFT formally, consider a broadcast system of nodes where a sender node needs to broadcast a message (value) to all the nodes in a peer-to-peer manner. 
At the beginning, the sender receives an input value $m$. The broadcast protocol must result in that at the end each node $i$ will output a value $m_i$. The sender and receivers may be honest or dishonest. This protocol achieves BFT if it satisfies two requirements:
	\begin{itemize}
	\item \underline{Consistency}: all honest nodes $i$ and $j$ must output the same value: $m_i = m_j$.
	\item \underline{Validity}: if the sender is honest, all honest nodes $i$ must output value $m_i = m$.
	\end{itemize}
A system can be consistent but not valid, when all honest nodes output the same value but this value is not the same as the sender's: $m_i = m_j \not \eq  m$. A system can be valid but not consistent, when the sender is dishonest and some honest nodes output different values: $m_i \neq m_j$ for some $i$, $j$. Thus, both requirements are needed.

Blockchain is a BFT system. To address inconsistencies due to the autonomous and independent working of blockchain nodes, the standard solution is for every node to agree on the consensus that the longest blockchain copy, the one with most blocks, is the globally correct version. Because the blockchain copies shorter than the correct blockchain are not used,  nodes  want to keep their copies as current as possible because otherwise they would waste efforts adding their blocks to a wrong blockchain.  As discussed in the previous subsection, nodes  frequently update their  blockchain copy to make sure its version is the latest (globally correct one). Consequently, even though at times some transaction may be recorded in different blockchain copies at different nodes,  different blocks may append to the same last block of the existing blockchain at different nodes, or different nodes may have different blockchain copies, eventually these nodes will have the same blockchain copy. 

But that is just  theory. If   consensus eventuality happens too late, the aforementioned inconsistencies will cause the system to perform incorrectly; for example, double spending can happen. Therefore, we need to 1) minimize the likelihood for inconsistencies to happen, and 2) minimize the time it takes to reach blockchain-consensus eventuality. Toward these, different  consensus mechanisms have been used for blockchain. Major among them are the methods of Practical Byzantine Fault Tolerance (PBFT) \cite{10.5555/296806.296824}, Proof of Work (PoW) \cite{bitcoin}, and Proof of Stake (PoS) \cite{10.1145/3132747.3132757}.

\section{The Bitcoin Network}
We present next the actual working of a real-world blockchain network: Bitcoin. It is a blockchain network to build a peer-to-peer digital cash system, where the name of the digital currency is bitcoin (BTC). It has a total supply of 21 million BTC to be minted over time according to a deterministic schedule such that all will have been minted in the year of 2140. Technically, it follows the general blockchain framework  described in the previous section. Specifically, it  adopts the UTXO model for the blockchain state  and the chain structure for the ledger. Newly arriving transactions will be put in a block to be appended to this chain. Any node can create blocks, and in that case it is called a ``miner'' and the process of creating a block is called ``mining''. The globally correct blockchain is chosen to be the longest one among all the local copies. We focus below on the key ideas and methods that are characteristic of Bitcoin implementation.

\subsection{Addresses}
To hold bitcoin, one needs to create a wallet. Each wallet corresponds to an address (the bitcoin address). When wallet $A$ is created, it is associated with a pair $(K^-_A, K^+_A)$ of 256-bit private key $K^-_A$ and 256-bit public key $K^+_A$  generated according to an asymmetric cryptography method called Elliptic Curve Cryptography (ECC) \cite{10.1007/3-540-39799-X_31,Koblitz1987,Hankerson2011}. Only the wallet owner knows the private key. The public key is publicly available. The address of wallet $A$ is a 160-bit hashed version of its public key $K^+_A$:
\[
A = RIPEMD160(SHA256(K^+_A)).
\]
This is one-way cryptographic hashing using RIPEDMD160 and SHA256 hash functions. 
Because only the owner has  the private key to unlock the public key, no one else can take ownership of a transaction that sends BTC to  her. For ease of human readability, Bitcoin addresses are  encoded as ``Base58Check'', which uses 58 characters (a base-58 number system) and a checksum, to produce a string like this example,  \emph{``1J7mdg5rbQyUHENYdx39WVWK7fsLpEoXZy"}.

\subsection{Elliptic Curve Cryptography}

\begin{figure}[t]
\centering
\includegraphics[width=\textwidth]{ 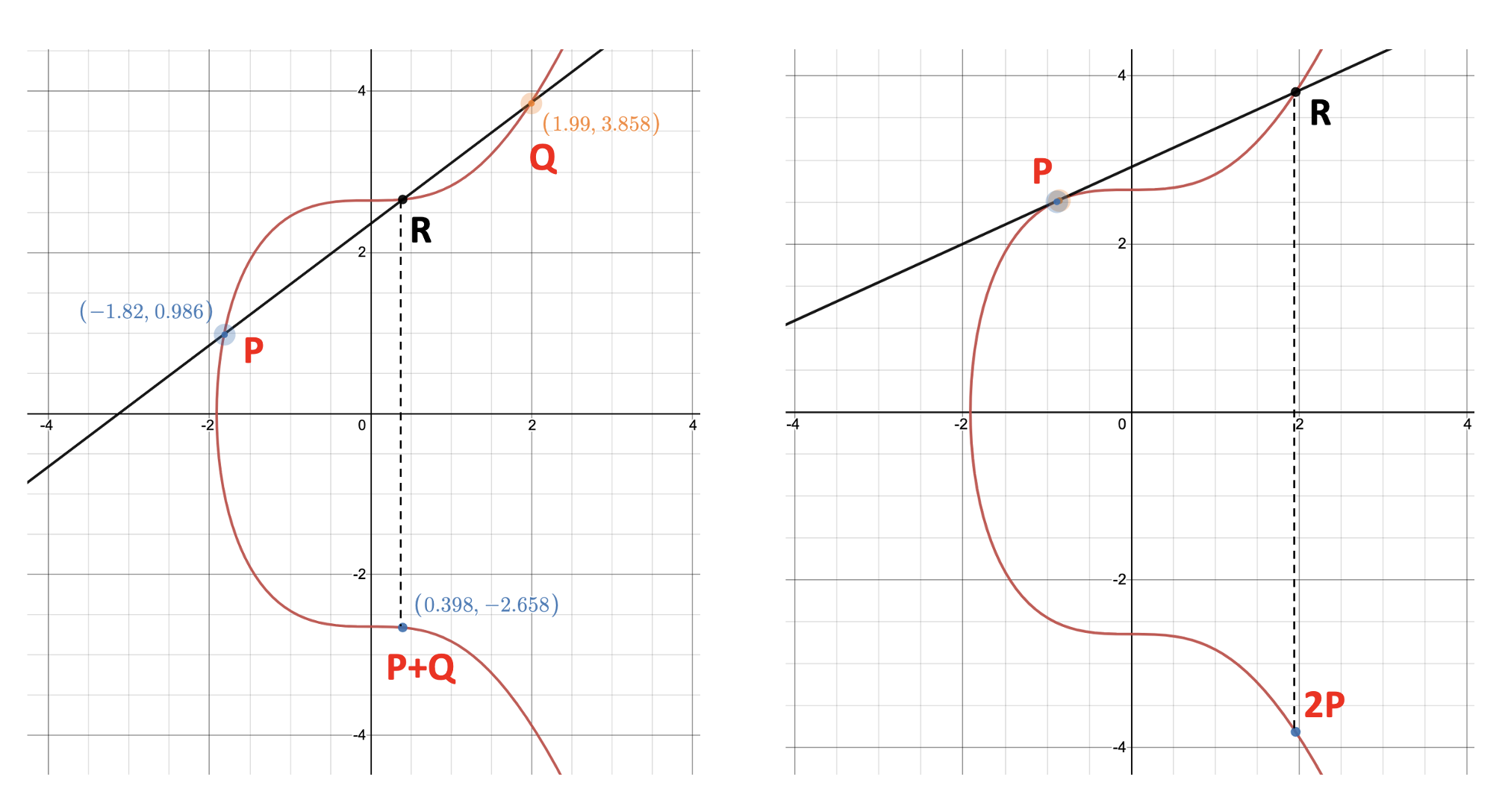}
\caption{Point addition on the elliptic curve ($y^2 = x^3 + 7$): (left) adding two different points; (right) adding two identical points.}
\label{fig:elliptic_add}
\end{figure}

The Elliptic Curve Cryptography (ECC) mentioned above is an approach to public-key cryptography based on the algebraic structure of elliptic curves over finite fields. The use of elliptic curves in cryptography was proposed in 1985 by Victor S. Miller \cite{10.1007/3-540-39799-X_31} and Neal I. Koblitz \cite{Koblitz1987} and became popular in 2004. For cryptographic purposes, an elliptic curve is a plane curve over a finite field (rather than the real numbers) with the following equation:

\newcommand{\Mod}[1]{\ (\mathrm{mod}\ #1)}

\[
y^{2} \equiv x^{3}+ax+b \Mod{p}
\]
The shape of the curve depends on the values given to $a$ and $b$. The size of the finite field is given by $p$, which defines the length of the keys we want to generate. The points on the curve are limited to integer coordinates within the square matrix of size $p \times p$ only. For example, the  curve in Figure \ref{fig:elliptic_add} is $y^{2} = x^{3}+7$ which is used in Bitcoin, and the  points in Figure \ref{fig:elliptic_integerpoint} are   integer points of $y^{2} \equiv x^{3}+7 \Mod{17}$.

On the elliptic curve, we define an algebraic operator on the points called ``point addition''. This operator allows  to ``add'' points to obtain a point on the curve, as follows (illustrated in Figure \ref{fig:elliptic_add}):
\begin{itemize}
\item \underline{Addition} $P + Q$: Draw the line $PQ$ and let $R$ be the point where $PQ$ cuts the curve. Point $P + Q$ is the mirrored point of $R$ over the x-axis. 
\item \underline{Double} $2P = P + P$: Draw the line tangent with the curve at point $P$ and let $R$ be the point where this line cuts the curve. Point $2P$ is the mirrored point of $R$ over the x-axis. 
\item \underline{Multiplication} $mP = P + P + ... + P$: This is the result of adding $P$ with itself $m$ times. 
\end{itemize}

\begin{figure}[t]
\centering
\includegraphics[width=0.7\textwidth]{ 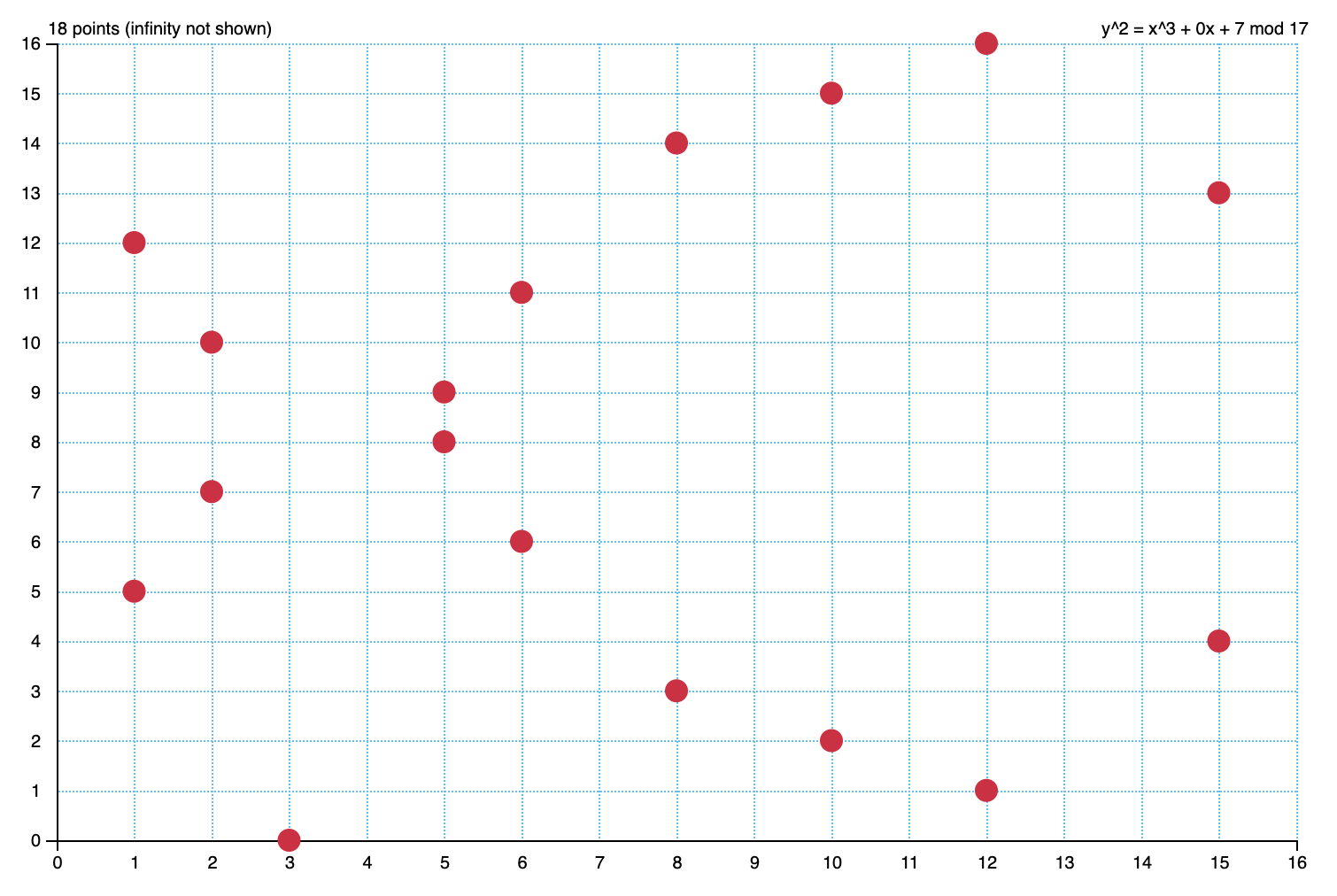}
\caption{The integer points of the elliptic curve on a finite field: $y^2 \equiv x^3 + 7 \Mod{17}$.}
\label{fig:elliptic_integerpoint}
\end{figure}

Despite its simplicity, a nice property of this operation on elliptic curves when applied on a finite field (i.e., all the points must be integer points in a finite square) is the hardness to compute the discrete ``logarithm'' $m$ such that  $mP = Q$  given points $P$ and $Q$. To date, no algorithm can reconstruct $m$ in time complexity faster than exhaustive search (having to try all possible values for $m$).   On the other hand, if some $m$ is given, it is easy to verify its correctness, that is to check whether $mP = Q$. For example, if $m = 16$, we need only $\log{m}=4$ point additions for verification: $2P$, $4P = 2(2P)$, $8P = 2(4P)$, and $16P = 2(8P)$; in comparison, to find the unknown $m$ in $mP=Q$ would need 16 point additions. 

Thanks to this property, ECC uses elliptic curves over finite fields to create a secret that only the private key holder is able to unlock. We can think of $Q$ as the public key and $m$ as the private key. The larger the key size, the larger the curve space, and the harder the problem is to solve. For example, Secp256k1 with equation $y^2=x^3+7$ and $p=2^{256} - 2^{32} - 2^9 - 2^8 - 2^7 - 2^6 - 2^4 - 1$  is  the ECC used by Bitcoin to implement its public key cryptography. All integer points on this curve are valid Bitcoin public keys.

\subsection{Transactions}
Bitcoin transactions are based on the UTXO model. A transaction by default is a transfer of BTC from a sender to one or more receivers. Every transaction must have a digital signature of the sender  who ``signs'' with her private key. This way, anyone who knows her public key can verify that the signature is valid and the transaction  indeed comes from that sender. Each output UTXO is destined for a receiving address. As we described earlier, each Bitcoin address is an encryption of its public key. Only the owner of that address has the corresponding private key to match. Hence, nobody but he can unlock the UTXO to use the fund.

\begin{figure}[t]
\centering
\includegraphics[width=0.82\textwidth]{ 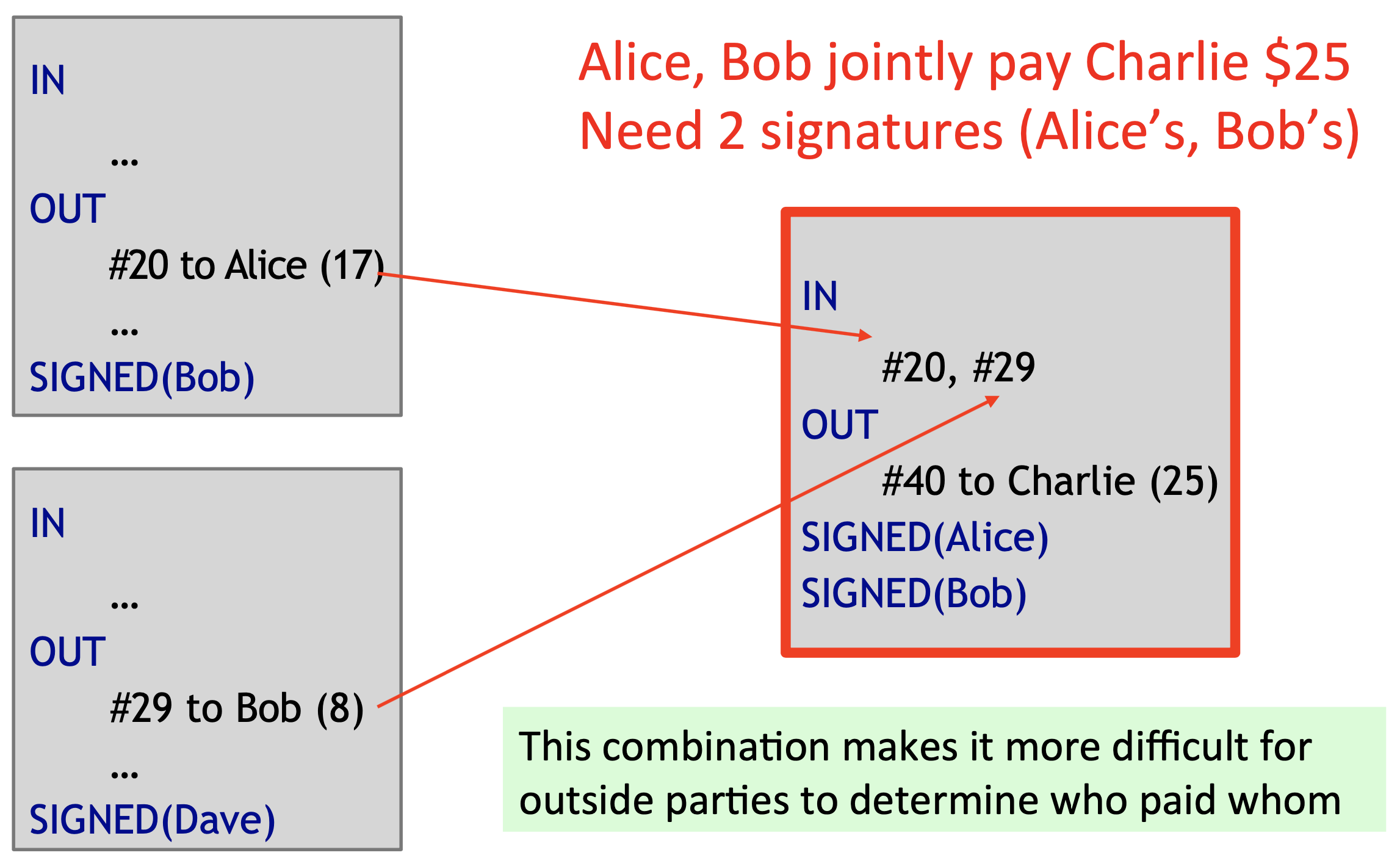}
\caption{Bitcoin joint payment: a transaction can use two or more input UTXO's that belong to different payers to collectively provide the fund to send. These multiple payers need to co-sign the transaction.}
\label{fig:utxo_jointpayment}
\end{figure}

A transaction can also be a joint payment which takes as input multiple UTXO transactions that belong to different addresses. For example, illustrated in Figure \ref{fig:utxo_jointpayment}, Alice and Bob jointly pay Charlie 25 BTC, where 17 BTC is funded by UTXO \#20 of Alice and 8 BTC is funded by UTXO \#29 of Bob. This joint transaction needs to be signed by both Alice and Bob. Joint payments make it more difficult for outside parties to determine who paid whom.

\subsubsection{Transaction Fee}
In a transaction, the input fund amount should be at least the output amount. The leftover is called the ``transaction fee'' to be sent to the miner who puts this transaction in a new block. Transaction fees are a way to incentivize miners to participate in Bitcoin. Rational miners prefer transactions that offer high transaction fees and so a transaction's sender  should choose a generous fee to increase its chance to be processed earlier. To determine the fee, the sender should consider the transaction size and the network traffic. A block can contain a maximum of 4 MB of data, thus limiting the number of transactions included. A larger transaction will take up more block data. Thus, larger transactions typically pay fees on a per-byte basis.

\subsubsection{Transaction Consolidation}

\begin{figure}[t]
\centering
\includegraphics[width=0.7\textwidth]{ 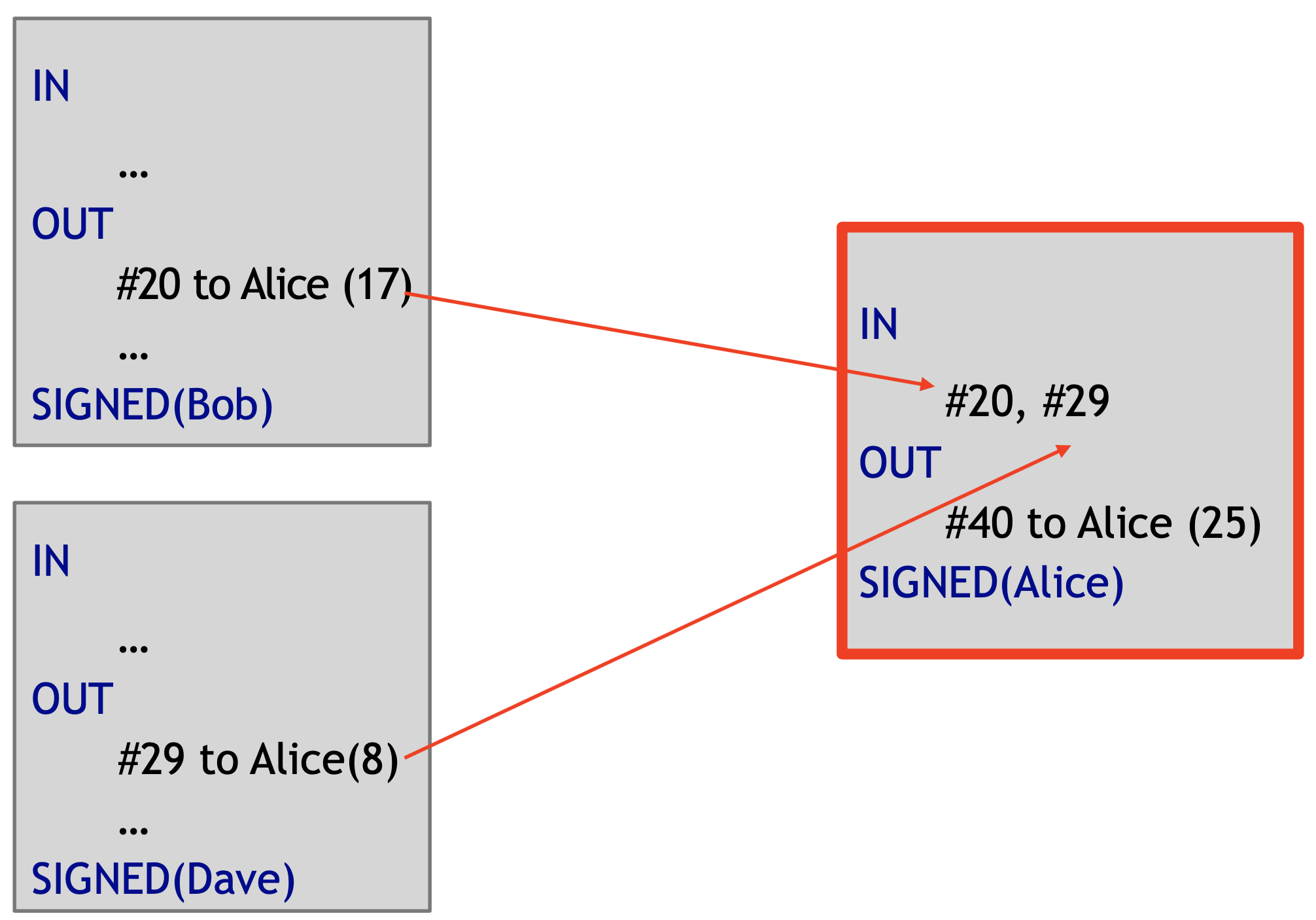}
\caption{Bitcoin transaction consolidation: an owner can create a transaction to consolidate the funds from many UTXO's he or she owns. }
\label{fig:utxo_2input}
\end{figure}

A consequence of Bitcoin's being a UTXO ledger is that one address may own many small UTXO transactions. As such, when this address makes a large payment out, it may need to include as input many UTXO's. Not only that the transaction size increases, but the transaction verification will be more expensive since it involves verifying many input UTXO's. For this large transaction to be included in a block, the sender should pay a high transaction fee. Therefore, it is a good idea for her to consolidate UTXO transactions if she owns too many of them. This can be done easily by creating a new UTXO transaction that consumes these existing UTXO transactions.  For example, as illustrated in Figure \ref{fig:utxo_2input}, Alice has funds in  UTXO  \#20 and UTXO \#29 and consolidates them by creating UTXO \#40. The decision for transaction consolidation is made at the application level by the wallet owner.

\subsubsection{Coinbase Transaction}
Transaction fees are not the only incentive for the miners. For each block that is successfully added to the blockchain, the  miner who created this block will receive a ``block reward''. As of March 2022, it is 6.25 BTC per block, which will  be halved automatically  after every 210,000 new blocks are added. To get the block reward, the corresponding miner, say Bob, inserts into the block a special transaction called the ``coinbase transaction'' that sends this 6.25 BTC to himself. This coinbase transaction has no input UTXO, meaning this amount will be minted by the network. If the block is validated and added to the blockchain,  all the transactions in this block, including Bob's coinbase transaction, are officially recorded, effectively sending the block reward to Bob. Coinbase transactions are the only way to mint bitcoin. Except the genesis bitcoin transfer, bitcoin is minted only by block mining, which is sent to the miners.

\subsection{Blocks}
Block creation is the main job of the miners. A miner pulls pending transactions from the mempool, typically selecting those with high transaction fees (because these fees will be paid to the miner) and put them into a block. This is called ``block mining''.  The very first block was added to Bitcoin network timestamped at 2009-01-03 13:15, called the genesis block, or block 0. It contains only one transaction, which is a coinbase transaction. This block is hardcoded in the Bitcoin client node software, so that when nodes join Bitcoin, they will always have the information about the genesis block.

\subsubsection{Block Structure}
A Bitcoin block has the following structure: 1) Block size (4 bytes): the size of the whole block in bytes; 2) Transaction count (variable size, 1 to 9 bytes): the number of transactions in the block; 3) Transactions (variable size): the list of transactions included in the block; and 4) Block header (80 bytes): important information useful for block creation and validation. The block header consists of the following fields:
	\begin{itemize}
	\item \underline{Version} (4 bytes): the version of the Bitcoin node software
	\item \underline{Previous hash} (32 bytes): the hash (ID) of the previous block
	\item \underline{Merkle root hash} (32 bytes): the hash value of the included transactions according to Merkle Tree
	\item \underline{Timestamp} (4 bytes): the block creation time in second (Unix epoch)
	\item \underline{Difficulty target} (4 bytes): a threshold number that is used for Bitcoin's Proof-of-Work algorithm
	\item \underline{Nonce} (4 bytes): a counter number that is used for Bitcoin's Proof-of-Work algorithm
	\end{itemize}

In Bitcoin, the ID of a block is a hash of its block header, not the whole block content. It is the  value resulted from hashing the block header  twice through the SHA256 algorithm. The block ID is not actually included inside the block’s data structure. Anyone can obtain this ID by applying double-SHA256 hashing on the block's header. 

\subsubsection{Merkle  Tree}

\begin{figure}[t]
\centering
\includegraphics[width=0.8\textwidth]{ 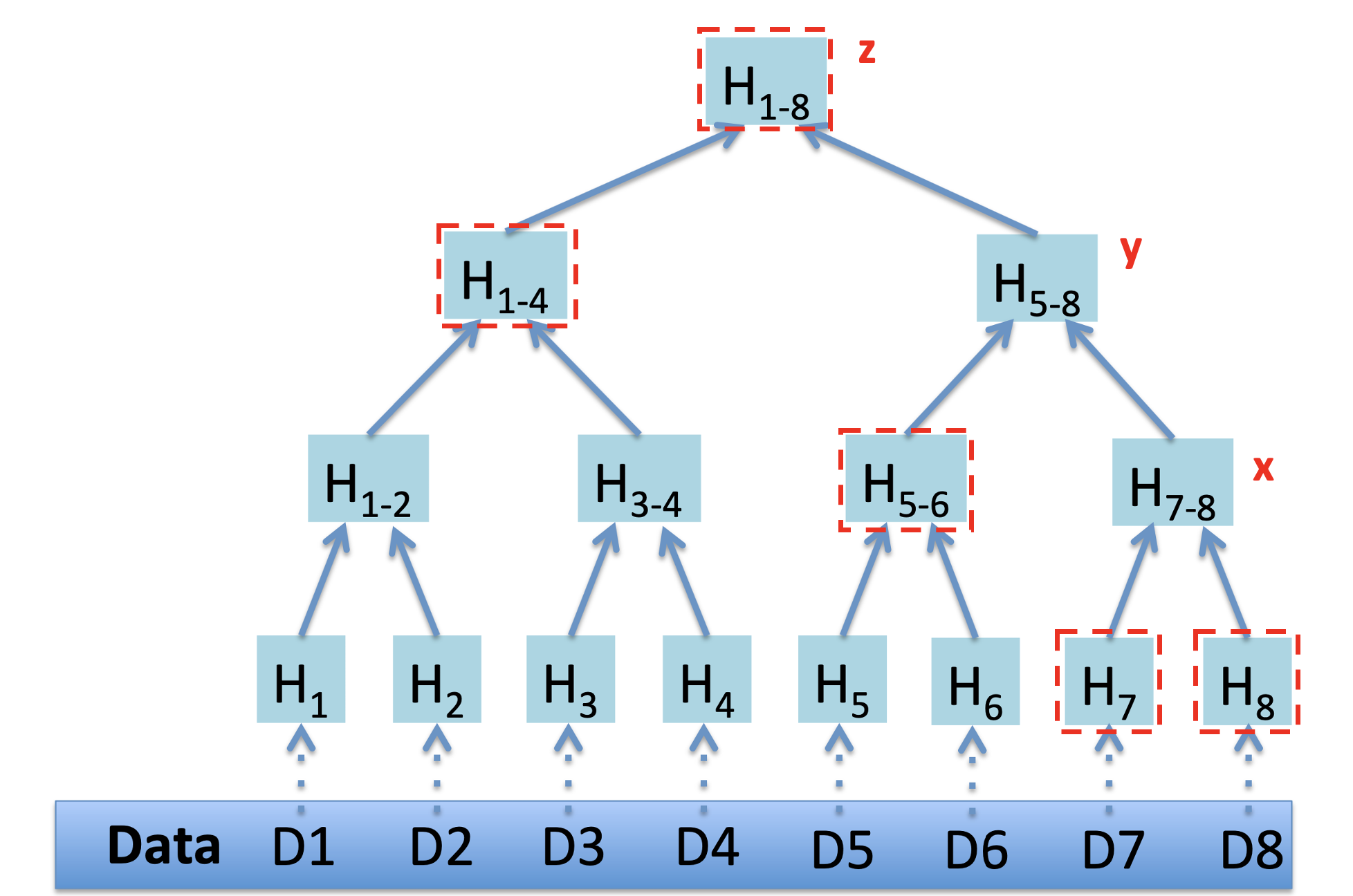}
\caption{Merkle tree: a binary tree where each internal node stores the hash value of the children's. }
\label{fig:merkle_tree}
\end{figure}

The transactions are organized in the block as a Merkle Tree \cite{10.5555/646752.704751}, a binary tree where each internal node stores the hash value of the children's values. Figure \ref{fig:merkle_tree} illustrates such a tree for Bitcoin, where there are eight transactions \{$D1$, $D2$, ..., $D8$\},  each  stored in a leaf node, internal node $H_{1-4}$ = $H( H_{1-2} \| H_{3-4})$, internal node $H_{1-2} = H(H_1 \| H_2)$, internal node $H_1 = H(D1)$, $H_2 = H(D2)$, etc. Bitcoin uses SHA2 for the hash function $H$.

The value at the tree root, e.g., $H_{1-8}$, is  the Merkle root hash stored in Bitcoin block header. There are two crucial properties. First, any change in the transaction data causes a change in the Merkle root hash. As such, if a block is altered, whether it is in the transaction data or the non-transaction part, the hash of the block will change and be detected. Second, it is fast to verify the existence of a transaction in the block. For example, to prove that transaction $D7$ is in the block, the prover only needs to provide four values as  evidence:  $H_7$, $H_{1-4}$, $H_{5-6}$, $H_8$. The verifier will compute the following
\begin{align*}
x &=&  H(H_7 \| H_8) \\
y &=& H\bigg(x \| H_{5-6}\bigg) = H\bigg(H(H_7 \| H_8) \| H_{5-6}\bigg) \\
z &=& H\bigg(y \| H_{1-4}\bigg) = H\bigg(H\bigg(H(H_7 \| H_8) \| H_{5-6}\bigg) \| H_{1-4}\bigg)
\end{align*}
and compare $z$ with $H_{1-8}$. Their equality means that transaction $D7$ is in the block. For a Merkle tree of $n$ transactions, it takes $\mathcal{O}(\log{n})$ time for this verification. It would take $\mathcal{O}(n)$ time if we naively store the transactions in an list-like structure.

A node in Bitcoin can be a non-miner node. It is only there to transact with  the network, uninterested in creating blocks to receive block reward. Because the block header provides sufficient information for verification and transacting purposes, the node needs  the block header only, not the full block content.  Since no actual transactions are stored, the storage requirement for such a node is modest (only 80 bytes, which can easily be placed in the memory). So is the communication cost to pull the blockchain copies from the neighbors (only pulling block headers).  

\subsubsection{Block ID Rule}

\begin{figure}[t]
\centering
\includegraphics[width=0.78\textwidth]{ 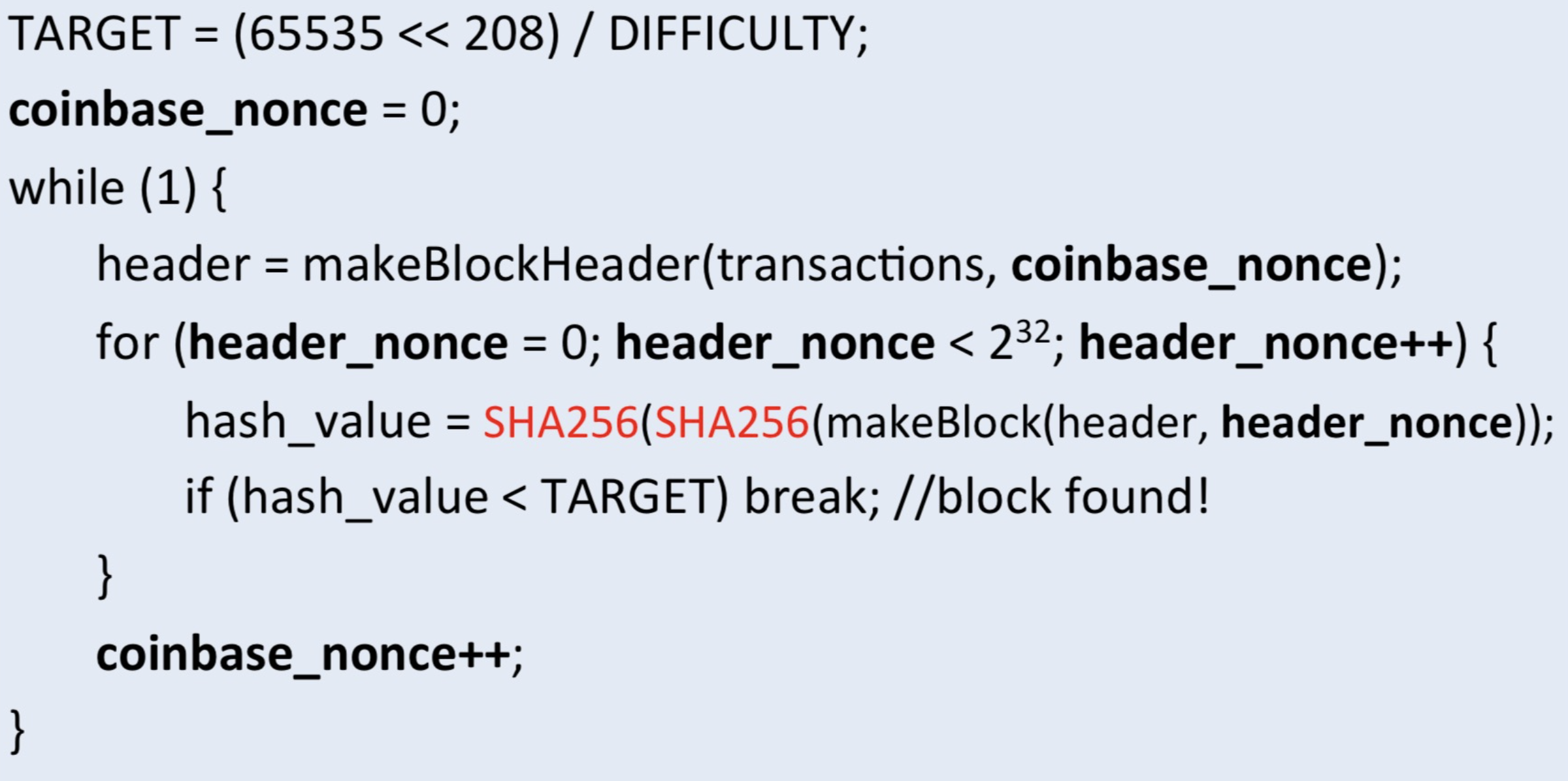}
\caption{Proof-of-Work mining algorithm in Bitcoin. }
\label{fig:pow_algorithm}
\end{figure}

After choosing transactions to put in a new block and organizing them into a Merkle tree, the miner needs to do a final critical step on the block before  submitting it to the network (by block broadcasting as we described in the general blockchain framework in Section \ref{sec:processatransaction}). This step is called ``mining'' or ``timestamping''. The block needs an ID and recall that the  ID is the hash value of the block header. In Bitcoin, the ID is not arbitrary, but follows a rule that it must be less than a target value determined by the \emph{difficulty} target value specified in the block header:
\begin{align}
\underbrace{SHA256(SHA256(block\_header))}_{ID} < \underbrace{(65535 << 208)/{difficulty}}_{TARGET}. 
\label{eq:difficulty}
\end{align}

The block header is known, except for two values that need to be filled: the $header\_nonce$ value and the $coinbase\_nonce$ value. $header\_nonce$ is the nonce attribute of the block header. $coinbase\_nonce$ is the value of the coinbase field in the coinbase transaction inserted by the miner to get the block reward. $coinbase\_nonce$ is set by the miner for flexible purposes.
We need to choose these two values such that the resulted  block ID satisfies the difficulty target according to Inquality (\ref{eq:difficulty}). 

\subsubsection{Proof-of-Work Mining}
To find a satisfactory block ID is not easy due to the double-SHA256 hashing. No algorithm is better than brute force, which takes $\mathcal{O}(2^{32})$: scanning all possible combinations of $header\_nonce$ value  and the $coinbase\_nonce$ value. A pseudocode of the mining algorithm is given in Figure \ref{fig:pow_algorithm}. The flexibility for the miner to set $coinbase\_nonce$ is important here because if we fixed it, we might not be able to find $header\_nonce$ among all the $2^{32}$ possible values such that the difficulty target is met. By allowing the miner to freely choose  $coinbase\_nonce$, eventually the miner will find a satisfactory block ID.

In Inequality (\ref{eq:difficulty}), if we increase $difficulty$, the value of $TARGET$ will decrease, making it more difficult to meet the inequality $ID < TARGET$. When Bitcoin started with the genesis block, the difficulty was set to $difficulty$ = 1, the easiest. Over the time, this target is  dynamically adjusted depending on the transaction traffic in the network. The Bitcoin node software updates this difficulty target such that the blockchain grows at a rate of one new block for every 10 minutes.

The idea of making miners solve the above computationally expensive inequality is to serve several purposes. If blocks are created too easily, since each newly created block is broadcast to the network, the communication cost would be very expensive. Worse, as the miners add blocks autonomously, many blocks simultaneously created by different nodes will be appended to the same last block of the existing blockchain (their local copy). This causes not only  severe inconsistency and  double  spending vulnerability, but also wasted efforts by  most miners due to the fact that only one block can append next to the existing global blockchain. Furthermore, malicious nodes can spam the network by creating many fake blocks and broadcasting them. 

Bitcoin resolves these problems by making the task of block creation difficult. A miner must spend some provable effort in order to create a block. This is analogous to real-world miners spending efforts to discover gold; hence the term ``Bitcoin miner/mining''. The found ID of the block is the proof, hence the term ``Proof-of-Work'' (PoW) always associated with Bitcoin. The challenge of finding a good ID satisfactory of Inequality (\ref{eq:difficulty}) is called the PoW problem. The PoW protocol also helps keep Bitcoin, as a currency, from inflation. Its slow minting and finite supply creates circulation scarcity, thus making its price valuable.

The concept of PoW is not new. It was proposed by Dwork and Naor \cite{10.5555/646757.705669} in 1992 to prevent email spamming. Every time you send an email, your computer must solve a computational puzzle. The recipient’s email program ignores your email if you do not attach the solution to the puzzle; if you do, the solution verification is quick. A similar idea was proposed in HashCash by Adam Back in 1997 and formally documented in 2002 \cite{hashcash} for anti-denial-of-service purposes. Bitcoin extended the PoW idea of HashCash.

\subsection{Mining Difficulty}

Achieving consensus in a permissionless environment is difficult due to Sybil attacks. As nodes communicate with one another in unauthenticated communication channels, a player can impersonate many others to outnumber the honest players and disrupt the consensus. This does not apply to a permissioned environment where the nodes are known to  the authority. The choice of the difficulty target for PoW is critical. The more difficult it is, the more robust Bitcoin is against Sybil attacks, as a bad player must pay a higher cost to harm the network. 

\subsubsection{Difficulty Setting\label{sec:difficultysetting}}

In Bitcoin, the PoW difficulty is set such that 1) the network is BFT (Byzantine Fault Tolerant) as long as more than half of the nodes are honest, say 51\%, and 2) on average only one block can be mined in each period of 10 minutes. To understand how they are related, we present a theoretical method below to find a good value for PoW difficulty (some derivations are similar to \cite{elaineshibook}). 

Let $n$ denote the number of blockchain nodes and $p$  the probability that a given node creates a block in a round (equal to 10-minutes in Bitcoin). We will set the difficulty target in Inequality (\ref{eq:difficulty}) to $TARGET = p2^m$ where $m$ is the hash bit length (256 bits for Bitcoin). Hence, $p$ indirectly represents the PoW difficulty. For example, if we set $p=1$, any given node will 100\%-certainly create a block, because Inequality (\ref{eq:difficulty}) is always satisfactory regardless of any block ID. However, that would result in $n$ blocks created, violating the 10-minute rule. Our goal is to find a good value for $p$.

The probability that no honest node creates a block in a round is $(1-p)^{0.51n}$. The probability to have a  block  created by some good node, hence a good block, in a round is
$1-(1-p)^{0.51n}$.
Consequently, the number of rounds it takes to mine a good block is
\[
\Lambda = \frac{1}{1-(1-p)^{0.51n}}.
\]
Let $\Delta$ be the worse-case network propagation time. It takes this much time for the mined block to reach all the honest nodes to be added to the good blockchain. An honest node would not produce the next block during this $\Delta$ period to make sure that the previous block must have reached all the nodes; else, the next block may be invalid (when validated at other network nodes before the previous block arrives there). Therefore, the block-mining efficiency is the ratio between the mining time to the actual time it takes for this block to be added to the blockchain:
\[
E = \frac{\Lambda}{\Lambda+\Delta} = \frac{ \frac{1}{1-(1-p)^{0.51n}}}{ \frac{1}{1-(1-p)^{0.51n}}+\Delta} = \frac{1}{1+\Delta(1-(1-p)^{0.51n})}.
\]

Let $q$ be the fraction of dishonest mining power, which is the total hashrate of all the dishonest nodes. We need the  hashrate of the honest nodes to exceed that of the dishonest, i.e., $(1-q) > q$. However, due to the efficiency $E$, the effective hashrate of the honest is $(1-q)E$, not $(1-q)$.  This block-mining efficiency $E$ does not apply to the dishonest nodes who can do whatever they want, for example sending block after block without considering the $\Delta$ delay. So, we should have $(1-q)E > q$. To make the blockchain even more secure, we introduce a parameter $\epsilon > 0$ arbitrarily small and make a more stringent requirement:
\begin{align}
\frac{effective~ hash~ rate~ of~ the ~honest}{hash ~rate~ of~ the~ dishonest} > 1+\epsilon.
\label{eq:epsilon}
\end{align}
The larger $\epsilon$, the more secure the network.
The left hand side is
\[
\frac{effective~ hash~ rate~ of~ the ~honest}{hash ~rate~ of~ the~ dishonest}  = \frac{(1-q)E}{q} = \frac{1-q}{q(1+\Delta(1-(1-p)^{0.51n}))},
\]
and so we require
\begin{align*}
&&\frac{1-q}{q(1+\Delta(1-(1-p)^{0.51n}))}
 >  1+\epsilon\\
&\Leftrightarrow& \frac{1-q}{q(1+\epsilon)} > 1+\Delta(1-(1-p)^{0.51n}) \\
&\Leftrightarrow& \frac{\frac{1-q}{q(1+\epsilon)}-1}{\Delta} > 1-(1-p)^{0.51n} \\
&\Leftrightarrow& (1-p)^{0.51n} > 1 - \frac{\frac{1-q}{q(1+\epsilon)}-1}{\Delta}  \\
&\Leftrightarrow& 1-p > \bigg(1 - \frac{\frac{1-q}{q(1+\epsilon)}-1}{\Delta} \bigg)^{1/(0.51n)},
\end{align*}
which leads to the following important inequality
\begin{align}
p < 1 - \bigg(1 - \frac{\frac{1-q}{q(1+\epsilon)}-1}{\Delta} \bigg)^{1/(0.51n)}.
\label{eq:p}
\end{align}
What this means is that we should choose $p$ to satisfy this inequality and the larger the gap between $p$ and this upper bound, the more secure the blockchain is.  We can choose a very small $p$ to make the mining difficult (because we set difficulty target to $TARGET = p2^m$), but doing so will slow down the transaction processing. A  good  $p$ is a reasonably high value still satisfying Inequality (\ref{eq:p}).

Inequality (\ref{eq:p}) also implies that given the same mining difficulty $p$, the blockchain  becomes less secure if the value of the right-hand side upper bound is smaller, because that creates a bigger risk for violating the inequality. The right-hand side will be smaller if the network delay $\Delta$ is longer or if the dishonest hashrate $q$ is faster. This explains why the security of a Bitcoin-like blockchain network is weakened in a slow network environment.

\subsubsection{Difficulty Adjustment}
Because each miner is solving the Proof of Work puzzle in parallel, on average the time taken for the first miner to solve it reduces inversely proportionally with the number of miners. At times when there are a lot of miners active on the network, the time to produce blocks will therefore be lower than when there are fewer miners active on the network. Since the number of miners on the Bitcoin network is going to change over time, unless some measure is employed, the block production time would also vary. In particular, over time if more and more miners joined the network, it would just keep decreasing. This is problematic as it could result in blocks being produced too fast, increasing the bandwidth requirements on the network, and also potentially result in more “forks”. 

\begin{figure}[t]
\centering
\includegraphics[width=\textwidth]{ 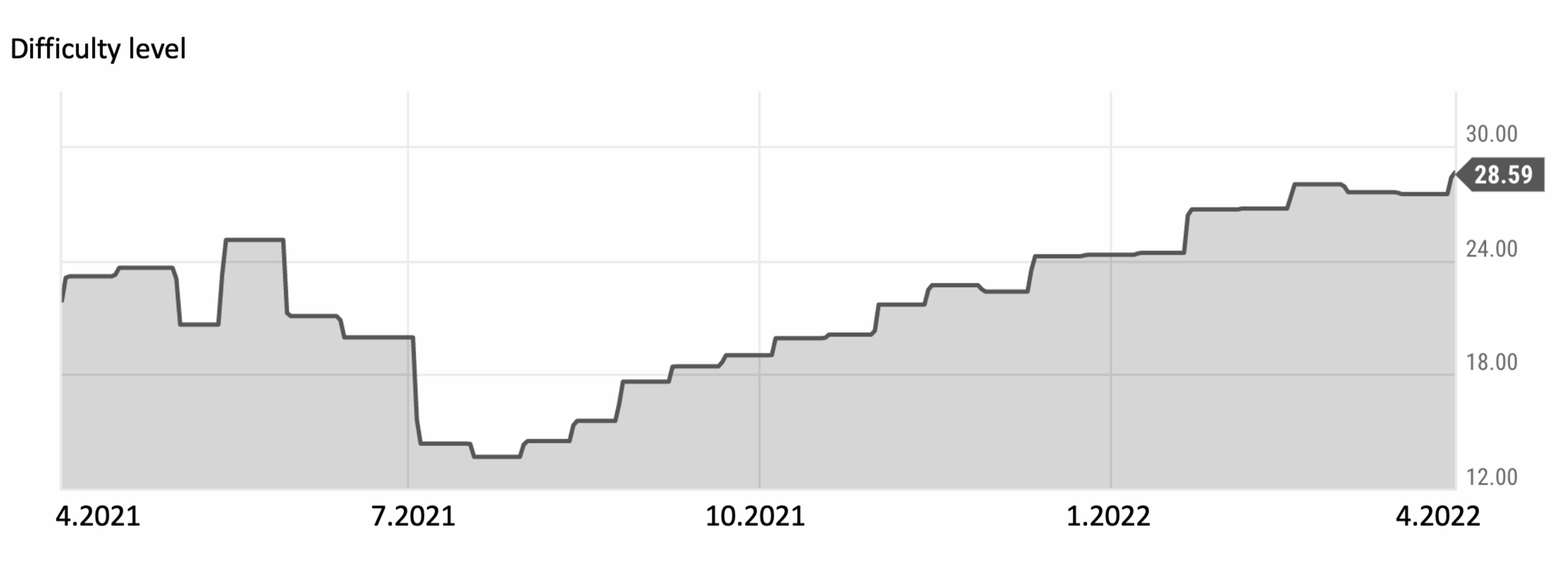}
\caption{Bitcoin PoW mining difficulty adjustment over the time as of April 1, 2022. The y-axis is measured in the number of `0' bits in the prefix of the difficulty target. }
\label{fig:difficulty}
\end{figure}

To prevent this, the difficulty level of block production is periodically adjusted in a decentralized manner in such a way as to ensure that on average a block is produced or mined once every 10 minutes.  Based on this 10-minute period, we can calculate that once every 2 weeks, the total number of blocks produced should be 0.1 (block / minute) x 60 (minutes / hour) x 24 (hours / day) x 7 (days / week) x 2 (weeks) = 2016 blocks. The protocol therefore adjusts the difficulty level after each epoch of 2016 blocks using the following equation: 
$D(n+1) =  \frac{2  D(n)}{T}$, Where $D(n)$ is the difficulty at epoch $n$, and $T$ is the time taken in weeks to produce the previous 2016 blocks. If this time is shorter than 2 weeks, then that implies that there must be too many miners on the network, and therefore the difficulty level should be increased; and on the contrary if this time is longer than 2 weeks, then that implies there are too few miners on the network and therefore the difficulty level should be decreased. 
Each miner independently computes the new difficulty and will only accept blocks that meet the difficulty that they computed. Figure \ref{fig:difficulty} shows how the Bitcon mining difficulty has changed during the past year (April 2021 - April 2022).

\subsection{Mining (Un)Fairness}

Bitcoin is generally fair in that if a node contributes more computing power, which is referred to ``hashrate'', it can solve the PoW problem more quickly, thus having a better chance to earn block reward. If a fast node A and a slow node B both want to add a block to the existing blockchain at the same time,  the block of A is  more likely to be born first (with satisfactory block ID) and will be added to the blockchain copies in the network before the block of B. Naturally, the PoW mechanism encourages nodes to upgrade computing power to receive more block reward. This healthy competition leads to a better Bitcoin network.

However, not all nodes are  good citizens. We have honest nodes who follow the protocol precisely and responsibly. There are selfish nodes who follow the protocol but do things to their benefit at the cost of other nodes. The remainder is the nodes who want to harm the network. To be equitably fair,  a node contributing a fraction of the network hashrate should have the same fraction of blocks accepted by the network (thus being rewarded).  For example, if a node contributes 20\% of total network hashrate, it should own  20\% of the blocks in the blockchain. In this aspect, Bitcoin can be very unfair.  We show below that Bitcoin can encounter a situation where the rate at which a good Bitcoin miner successfully adds a block to the blockchain is much lower than its hashrate contribution.

 \subsubsection{Selfish Miner Attack}
First, let us explain how a selfish miner A can abuse the network. Each time A has created a block $a$, it does not broadcast the block to the network right away. Instead, A waits on the event of  receiving a new block from another miner, say block $b$ of miner B. When this happens, A will immediately broadcast its block $a$ to the network, but ignore block $b$ of miner B (not forwarding it further).   If we generalize this strategy such that A represents the pool of selfish miners and B the pool of honest miners, the blocks belonging to the selfish will be faster to reach the blockchain nodes than the honest's blocks. This is because while A honestly forwards every block, including B's, immediately upon its arrival, B ignores A's blocks. As a result, not only B receives more block reward, but also a lot of A's efforts are wasted.

Bitcoin's vulnerability to Selfish Mining Attacks was investigated by Eyal and Sirer in \cite{10.1145/3212998}. It is shown that selfish miners can collude to obtain a revenue larger than their fair share. This attack is potentially serious in that rational miners may prefer joining the selfish miners leading to a 51\% majority, thus destroying the decentralization of Bitcoin.

\subsubsection{Unfairness Severity}
Next, let us see how unfair Bitcoin can be from a theoretical view given dishonest behaviors. Consider the formulation in Subsection \ref{sec:difficultysetting}.  It is shown  in \cite{elaineshibook} that, in an honest node's blockchain,  the honest block fraction is approximately $\mu = 1-\frac{1}{1+\epsilon}$. Assume a blockchain network with zero propagation delay $\Delta=0$, which is ideal for honest nodes because the block creation efficiency is $E=100\%$. The honest hashrate is therefore $(1-q)E=1-q$.  Recall  $q$ as the total hashrate fraction of the dishonest nodes. Inequality (\ref{eq:epsilon}) becomes
\[
\frac{1-q}{q}  > 1+\epsilon.
\]
As a result, we have
\begin{align*}
\mu = 1-\frac{1}{1+\epsilon} < 1 - \frac{q}{1-q} = \frac{1-2q}{1-q} 
\end{align*}
which is smaller than the  honest hashrate fraction $(1-q)$.

What this means is that even in a network ideal for the honest nodes, the rate at which they can add blocks to the blockchain, $\mu$, is lower than the hashrate they contribute to the network, $(1-q)$. For example, when 51\% of the network is honest (i.e., $q=49\%$), they own a fraction $\mu < \frac{1-2q}{1-q} = 3.9$\% of the blocks on the blockchain. The dishonest coalition   with only 49\% hash power owns 96\% of the block creation.

So, Bitcoin mining may be unfair when it comes to block reward.  Despite this risk in theory, one may argue that it is unlikely or of little impact in practice. Miners have ideological considerations and incentives to keep the network decentralized. If a coalition grows so big to be a concern to the rest of the network, people may leave Bitcoin due to the lack of decentralization; the coalition would not benefit, of course.

\subsection{Block Finality}
A malicious miner makes a payment, then secretively creates a second conflicting transaction using the same UTXO input in a new block, that allows him to recover the fund. This is an example of the Double Spending problem. This is feasible if this miner controls more than $50\%$ hashrate of the whole network, mining faster than the rest of the network combined. Therefore, his local chain is the longest among all local copies and will be accepted by the network as the consensus for the globally correct blockchain.

Even when the bad minor has less than 50\% hashrate as in most cases, there is still a non-zero chance that the bad miner can grow the longest blockchain. Although this can only last   for a short period of time, double spending is not impossible. To minimize this risk, when somebody pays a merchant to buy something, the merchant should wait some time to make sure the money is in before delivery. In Bitcoin, the wait is for 6 block confirmations, i.e., 6 blocks to be added after the block containing the payment transaction. 

Why 6 block confirmations is enough? Consider a miner $A$ with a fraction $p$ of the total hashrate  
and a miner $B$  with a smaller fraction $q=1-p < 1/2$. We are interested in computing the probability that $B$'s blockchain will be longer than $A$'s after $A$ adds $k$ blocks if both nodes start at the same time. This is similar to a race of two players in a Gambler Ruin problem. 
In this game, block creations form a sequence of independent Bernoulli trials. Each trial is the creation of a block which has two potential outcomes: ``success'' means that the block is created by $A$ and ``failure'' if the block is created by $B$. 

 We observe this sequence until $A$ has created $k$ blocks (i.e., $k$ successes). The number of blocks $B$ created is a Negative Binomial random number, $X \sim NB(k, p)$, which has the following probability mass function:
\[
P(X=i) = {i+k-1 \choose i} (1-p)^ip^k.
\]

During the time that $A$ has added $k$ blocks, the probability that $B$ has created more than $k$ blocks, hence winning the race outright, is
\begin{align}
P(X > k)  =  \sum_{i>k} P(X=i) = \sum_{i>k} {i+k-1 \choose i} q^ip^k.
\end{align}
In the case that $B$ has created less than or equal to $k$ blocks, i.e., $i \le k$, $B$ will be behind by $(k-i)$ blocks and still has  a probability $(q/p)^{k-i}$ to catch up with $A$ and thus win. Summing up these   probabilities will lead to the probability that $B$ will win the race:
\begin{align}
P(k) 
= P(X > k) + \sum_{i=0}^k (q/p)^{k-i}P(X = i)\\
= \sum_{i>k} {i+k-1 \choose i} q^ip^k + \sum_{i=0}^k (q/p)^{k-i} {i+k-1 \choose i} q^ip^k\\
= \bigg ( 1 - \sum_{i=0}^k {i+k-1 \choose i} q^ip^k \bigg)+ \sum_{i=0}^k  {i+k-1 \choose i} q^kp^i\\
= 1 -  \sum_{i=0}^k {i+k-1 \choose i} (q^ip^k - q^kp^i).
\end{align}
This probability converges exponentially to zero as $k$ increases. Grunspan and Perez-Marco   \cite{DBLP:journals/corr/abs-2003-00001} provides a closed form for this probability
\[
P(k) = I_{4pq}(k, \frac{1}{2})
\]
where $I(.)$ is the regularized incomplete beta function:
\begin{align*}
I_x(a, b) = \frac{\Gamma(a+b)}{\Gamma(a)\Gamma(b)} \int_0^x t^{a-1}(1-t)^{b-1}dt.
\end{align*}

For Bitcoin, it is recommended that we wait for $k= 6$ block confirmations before assuming that the transaction is final, which is enough for $P(k)$ to be extremely small. For example, $P(6)=0.0005914$ for $q=0.1$. A block becomes ``final'', hence the blockchain up to this block is considered ``finality'', if it is followed by this many block confirmations. Thus, Bitcoin finality  is not instant. Instead, it is guaranteed asymptotically.

\section{Smart Contract Blockchains}

Bitcoin is an example of an application-specific blockchain network where the only application logic is to serve digital payments. Although it allows for some limited programmability, it does not provide arbitrary programmability. As many applications in the real world, not necessarily financial, can benefit from blockchain technology, having a dedicated blockchain network for each individual application is not realistic.

This is the motivation for Ethereum, the first blockchain network created by Vitalik Buterin et al. in 2012 \cite{ethereum} to be a universal blockchain computer that can run  applications of arbitrary purposes. To develop such applications, developers write computer programs called ``smart contracts". Ethereum, therefore, is said to be a smart contract blockchain. Other public smart contract blockchain networks include Algorand \cite{DBLP:journals/tcs/ChenM19}, Tezos \cite{tezos}, and Solana \cite{solana}.

\subsection{Smart Contract}
Smart contracts are written using a high-level programming language (e.g., Solidity, Viper, Flint, Bamboo). Solidity is the most popular language for smart contract networks. It is Turing-complete, meaning that  it can simulate any computation. In contrast, Script, the programming language of Bitcoin, is not Turing-complete. Script is thus very light and suitable for Bitcoin. Bitcoin does not need a universal language because  digital currency is the only purpose of Bitcoin. 

Compared to Bitcoin, a smart contract blockchain has an additional layer of functioning because of the smart contract capability. When a smart contract is deployed, it is submitted as a transaction to the blockchain network to run on every node. 
Each node needs a  runtime environment to execute the bytecode of the smart contract. On Ethereum, this is called the Ethereum Virtual Machine (EVM), a powerful sandboxed virtual stack embedded with each full Ethereum node.  EVM is where all Ethereum accounts and smart contracts live. It maintains the consensus for the blockchain. While the smart contract language used in Ethereum is Solidity which is Turing-complete, EVM is a quasi-Turing complete machine. Quasi, because EVM can theoretically run every smart contract but its execution will stop and be reverted if exceeding the resource allocation limit specified by the deployer.
 
As an analogy, let us compare running a computer program on a single computer versus the Ethereum blockchain computer. In the former case, all the state of the program is stored on the computer which is the only point of contact for the user and if this computer fails, all the state will be lost. In the blockchain case, the computer program is deployed and runs simultaneously on all computer nodes of the blockchain; these nodes independently and autonomously keep track of the program's state. A user can interact with the program using any node. If a node fails, the program is still running on the other nodes. The consensus mechanism of the blockchain ensures that the states on all the nodes are identical. 

The source codes of deployed smart contracts are visible to the public. Therefore, there is nothing to hide in the working of a smart contract and people can be assured that it will work as programmed. That said, in certain cases, a complex smart contract may contain bugs and other security holes that are not easily seen; to fix them is a headache after the application is already deployed with many users; note that the blockchain is immutable. Therefore, a professional project should have its smart contracts certified by reputable smart contract auditors.

\subsection{Token Creation}
Bitcoin (BTC) is the only native token (digital currency) of the Bitcoin blockchain network. Its users transact with each other (paying one another) using BTC.  A smart contract network also has a native token (e.g., ETH for Ethereum), whose main use is for the users to deploy smart contracts and interact with them. To deploy a smart contract on Ethereum, one must pay a certain amount in ETH. This amount depends on the computational complexity of the contract. Besides ETH, many secondary tokens can be created to serve different applications. For example, one  can build a loyalty application on top of Ethereum and implement the loyalty point as a token, or a country's government can issue a central bank digital currency (CBDC) as a token on top of Ethereum. 

\begin{figure}[t]
\centering
\includegraphics[width=\textwidth]{ 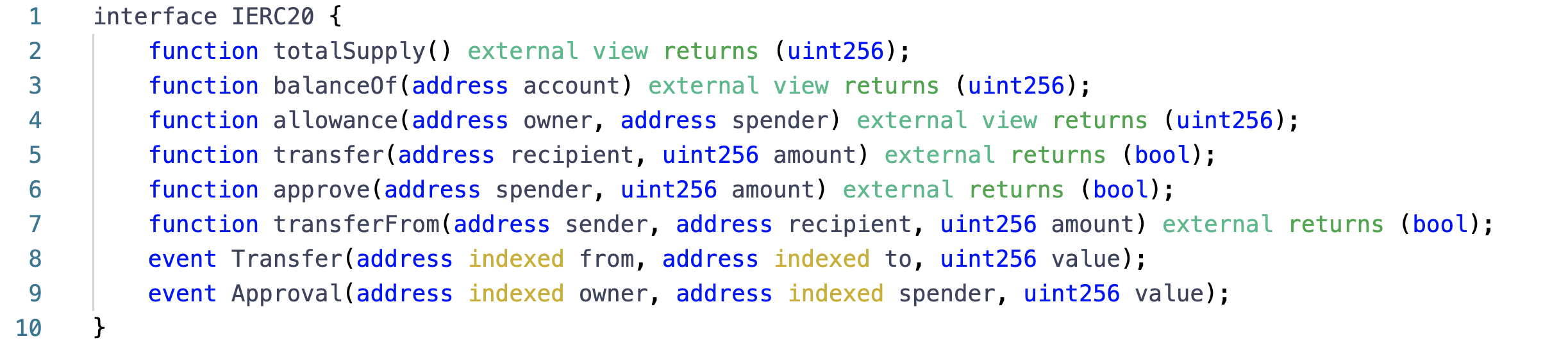}
\caption{The IERC20 interface for ERC-20 tokens.  ERC-20 tokens must implement these functions. }
\label{fig:ierc20}
\end{figure}

A token is implemented in the form of a simple smart contract. If the token is meant to be a kind of digital currency, this contract stores the token balance information for each account (blockchain address) and includes essential functions to enable a sender to transfer tokens to a receiver (needed for a real-world payment transaction), a spender to transfer tokens on behalf of their owner to a receiver (useful for a trading exchange or a bank to transfer money from someone's account to a payee, of course with permission only), or in many cases, mint new and burn existing tokens (useful for a government to cope with inflation crises). Figure \ref{fig:ierc20} shows the interface for token smart contracts in Ethereum with six required functions. Events can be emitted from a smart contract  so that front-end applications can watch and be instantly notified of their happening.

\begin{figure}[t]
\centering
\includegraphics[width=\textwidth]{ 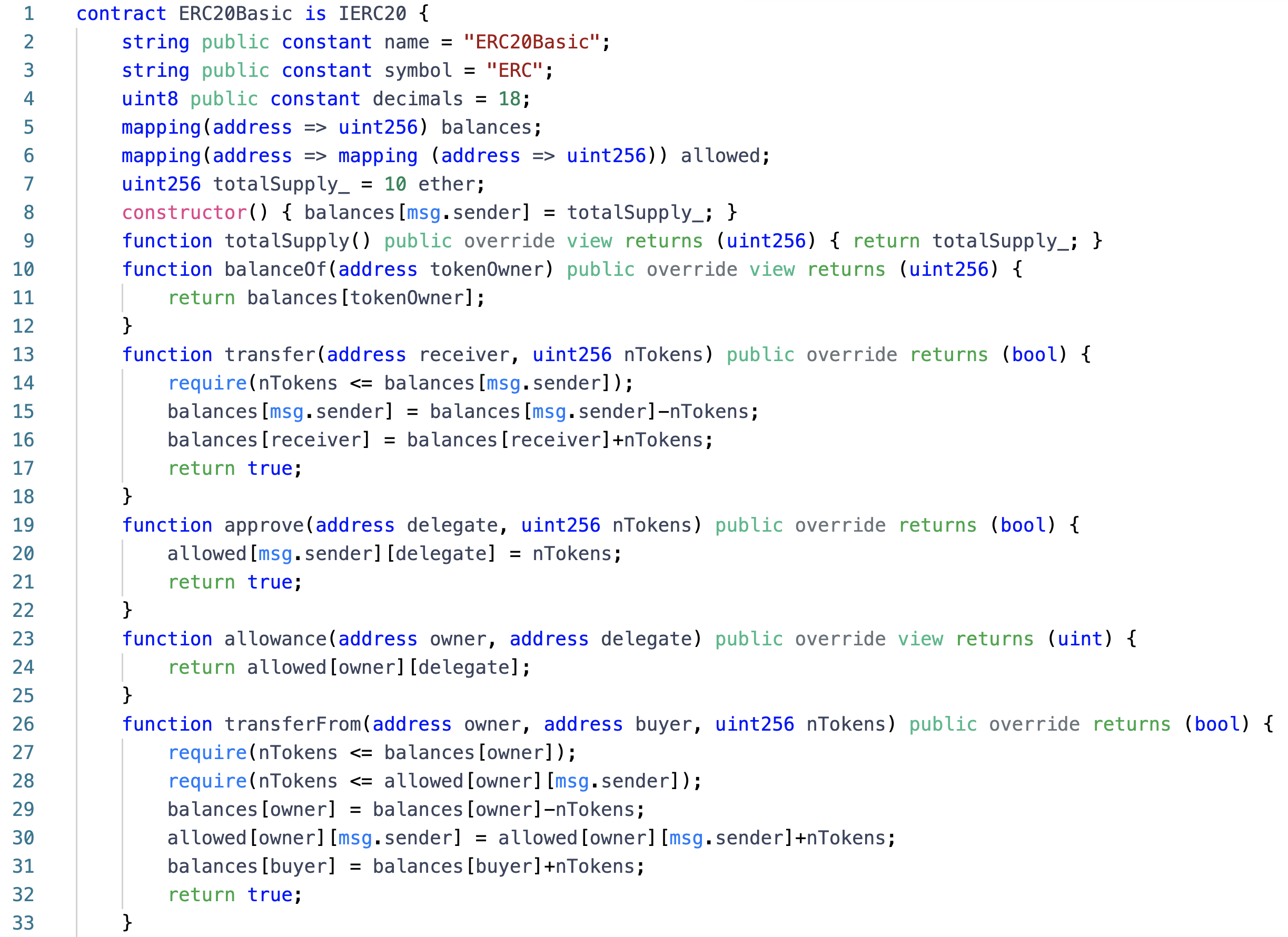}
\caption{A basic smart contract of an ERC-20 token implementing IERC20 interface. }
\label{fig:erc20}
\end{figure}

For ease of token creation, several token standards have been defined and template smart contracts created. The first standards were defined for the Ethereum network and their counterparts later followed for other smart contract networks. For Ethereum, ERC-20 is the standard for fungible tokens, ERC-721 for non-fungible tokens (NFT), and ERC-1155 for generic multi-tokens (one that can represent a  fungible token or a non-fungible token or a multiple of them). For example, ERC-20 is used for implementing a  cryptocurrency, ERC-721 for digitally representing a physical asset uniquely and non-duplicatable as an NFT, and ERC-1155 for digitally representing equity shares of a company. A basic Solidity smart contract implementation of ERC-20 is shown in Figure \ref{fig:erc20}.

\subsection{Transaction Processing}
Since a transaction may involve interacting with a smart contract by calling a function in the smart contract, the processing is not simply a verification such as checking fund availability. Let us explain this for the case of Ethereum. The Ethereum blockchain adopts the account-based state model; its state consists of a set of accounts (blockchain addresses) and their corresponding information. There are two types of accounts:
\begin{itemize}
\item \underline{Externally-owned account}: one that is owned by a normal user (like a bitcoin account). The state  for an external account is its ETH balance.  
\item \underline{Contract account}: one that represents a deployed smart contract. The state for a contract account consists of its ETH balance,  contract code, and a storage area to save the run-time state of the contract. 
\end{itemize}

The Ethereum blockchain protocol is essentially the same as that in the blockchain framework we presented in Subsection \ref{sec:processatransaction}, the main difference being in transaction processing and block validation steps.

A transaction is a transfer of asset/value and optional data from a sender to a receiver. Specifically, it has a sender who initiates the transaction, a receiver who receives the transaction, a value (amount of tokens) to be transferred from the sender to the receiver, and a data part if the receiver is a contract account.  If the transaction is received by a contract account, the corresponding contract code will be executed, taking as input the data included in the transaction. Ethereum introduces a concept called ``gas fee'' to  represent how much ETH the transaction will pay the miner. A transaction contains two values, $startgas$ and $gasprice$. 
\begin{itemize}
\item The $startgas$ value represents the maximum number of computational steps the transaction execution is allowed to take. The sender should have an idea as to how complex the transaction is and determines this value properly. If the miner takes more steps than this threshold allows, the transaction will halt and be reverted. 
\item The $gasprice$ value is the ETH fee the sender will pay the miner per computational step. To expedite the processing, the sender should increase $gasprice$ so that the miner would include the transaction in the next block.
\end{itemize}

Upon a transaction, the \underline{state transition} happens as follows:
\begin{enumerate}
\item Check if the transaction is well-formed and valid. Else, terminate.
\item Set transaction fee to $startgas \times gasprice$. Subtract this fee from the sender's balance. If the balance is not sufficient, terminate.
\item Initialize $GAS = startgas$, minus a certain quantity of gas per byte to pay for the byte count in the transaction.
\item Transfer the value specified in the transaction from the sender's balance to the receiver's balance.
	\begin{itemize}
	\item If the receiver does not exist, create it.
	\item If the receiver exists and is a contract account, run the contract code either to completion or until the execution runs out of $GAS$.
	\end{itemize}
\item If this transfer fails: revert all state changes except the payment of the fees, and add the fees to the miner's account.
\item Else, refund the remaining $GAS$ to the sender, and send the fees paid for gas consumed to the miner.
\end{enumerate}

\subsection{Block Validation}
In Ethereum, a block contains a list of transactions and the blockchain state obtained by applying these transactions to the previous state (stored in the previous block). The creation of a new block requires PoW mining similar to Bitcoin (although, Ethereum is transitioning to Proof-of-Stake consensus in version 2.0). The validation of a block requires  checking the cryptographic link with the previous block as usual, but in addition, it has to replay the running of all the transactions in the block. Specifically, a miner validates a new block  as follows:
\begin{enumerate}
\item Verify that the previous block referenced exists and is valid.
\item Verify that the timestamp of the new block is greater than that of the previous block and less than 15 minutes into the future.
\item Verify that the new block's ID, difficulty target, transaction Merkle root are valid.
\item Starting from the previous blockchain state (stored in the previous block):
	\begin{itemize}
	\item Run a sequence of state transitions as a result of applying all the transactions in the new block, one by one.
	\item If any such transaction replay fails or if the total gas consumed exceeds the limit, terminate.
	\end{itemize}
\item If the Merkle tree root of the final state in the above step equals that stored in the new block, then the new block is valid. Else, invalid.
\end{enumerate}

\subsection{Contract Interoperability}

\begin{figure}[t]
\centering
\includegraphics[width=0.73\textwidth]{ 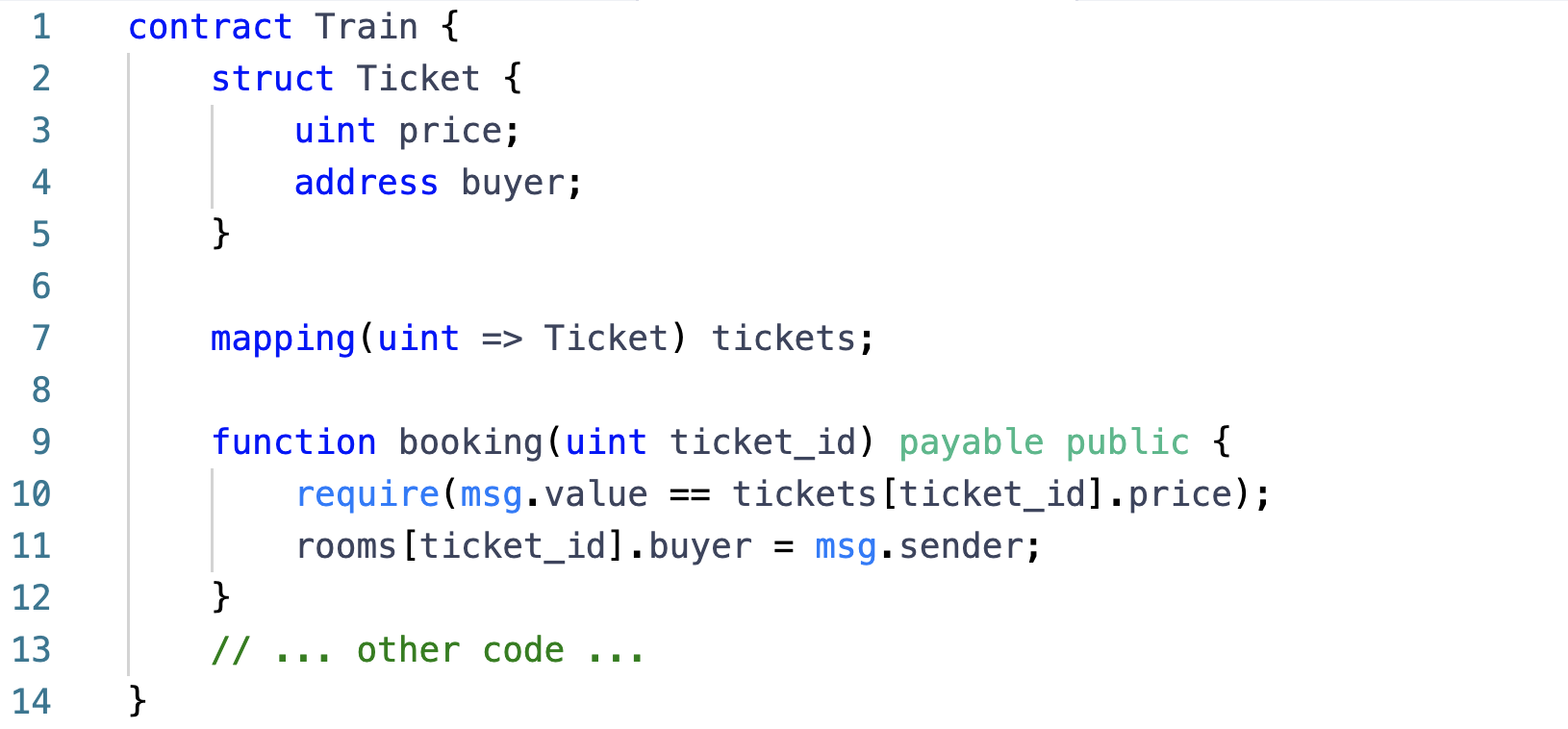}
\caption{The Train smart contract: this contract will be called later by the Booking smart contract. }
\label{fig:train}
\end{figure}
\begin{figure}[t]
\centering
\includegraphics[width=0.73\textwidth]{ 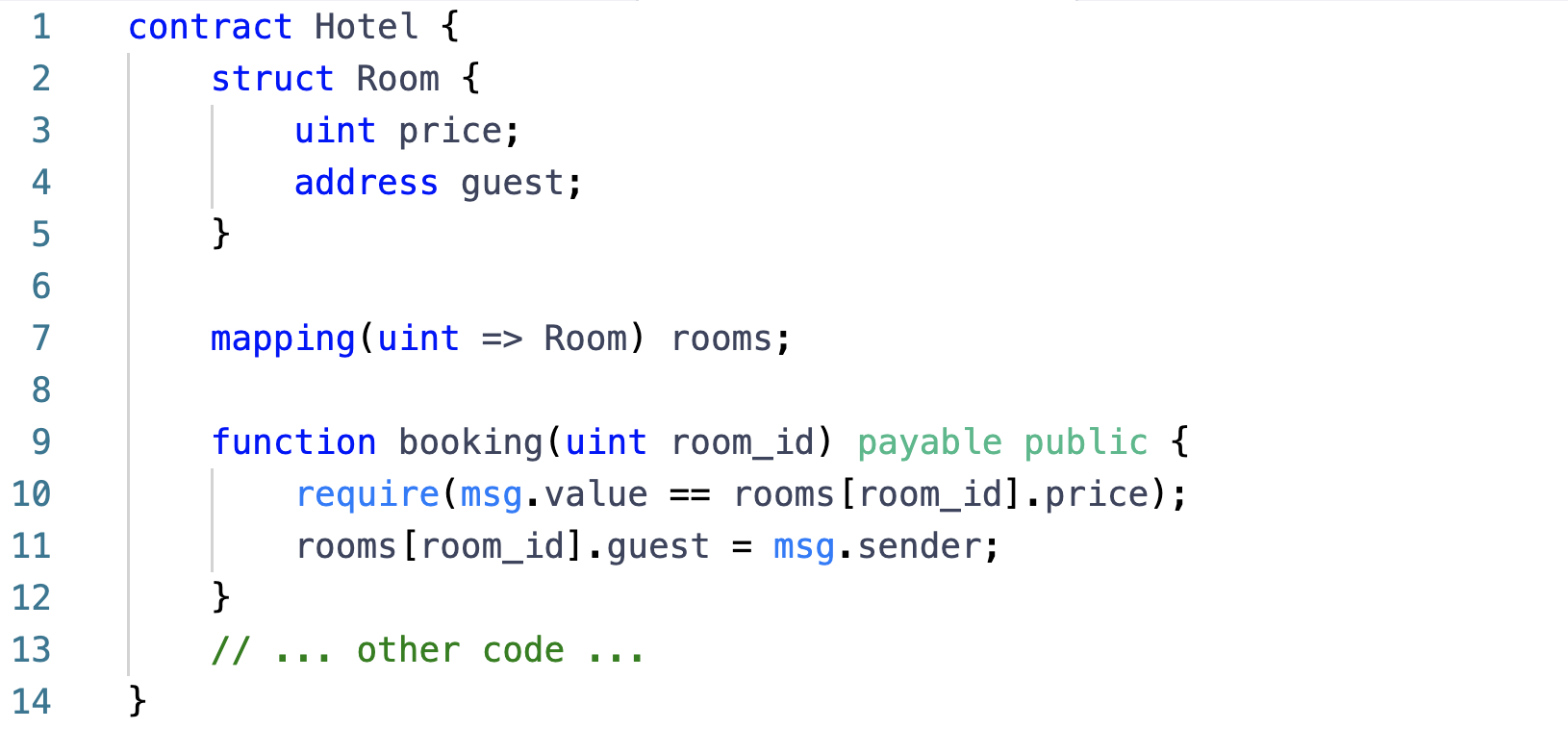}
\caption{The Hotel smart contract: this contract will be called later by the Booking smart contract. }
\label{fig:hotel}
\end{figure}

\begin{figure}[t]
\centering
\includegraphics[width=0.73\textwidth]{ 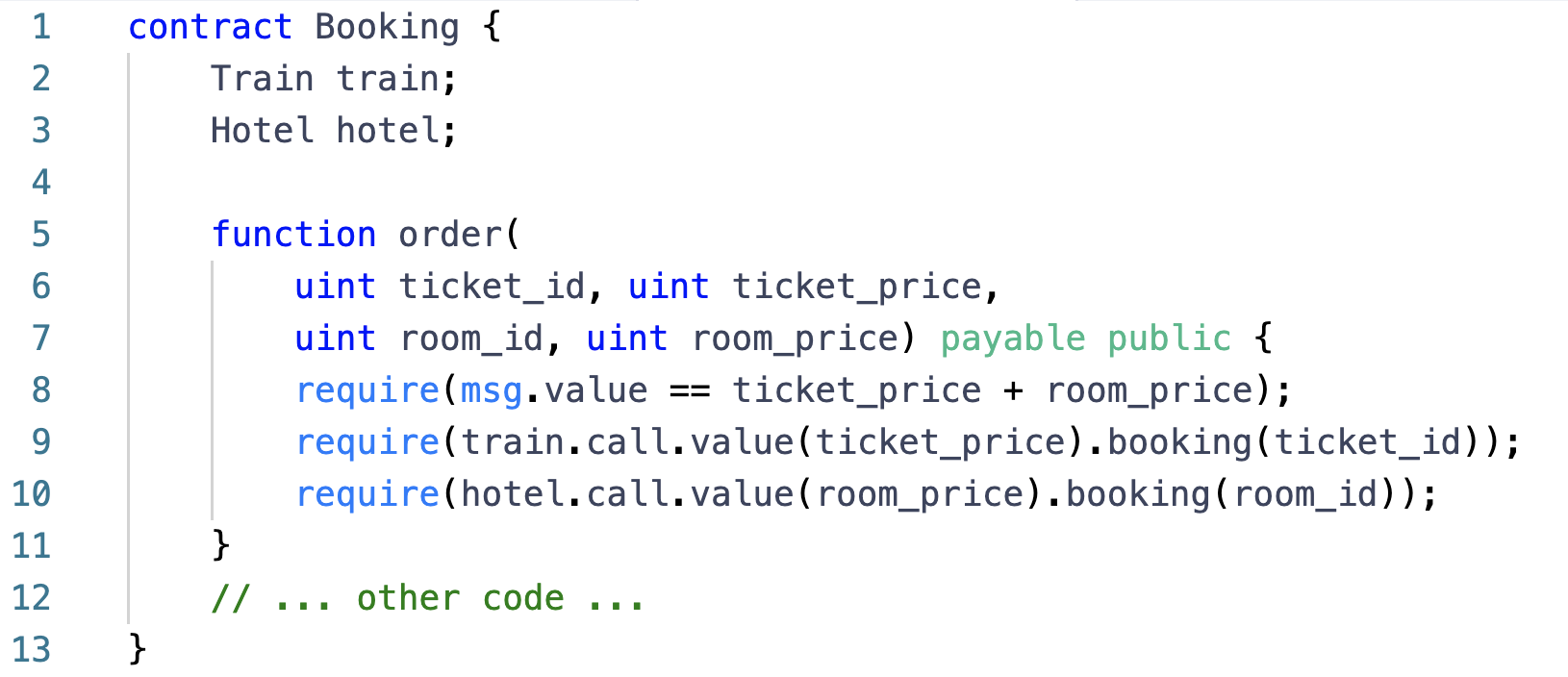}
\caption{The Booking smart contract: this contract calls the Train contract and Hotel contract. }
\label{fig:booking}
\end{figure}

In a smart contract blockchain network, one can call another contract inside a contract. For example, consider a Travel Booking application: allow people to purchase a train ticket and reserve a hotel such that each booking is atomic - either both reservations succeed or neither do. This is referred to as the ``train-and-hotel'' problem popular as a case study in Ethereum research community\footnote{This problem is described on this page: https://eth.wiki/sharding/Sharding-FAQs.}.

We have three smart contracts: 
\begin{itemize}
\item \underline{Train smart contract} (Figure \ref{fig:train}): keep status of all train bookings. A user can book a train ticket by calling the \texttt{booking()} function of this contract.
\item  \underline{Hotel smart contract} (Figure \ref{fig:hotel}): keep status of all hotel bookings. A user can book a hotel room by calling the \texttt{booking()} function of this contract.
\item \underline{Booking smart contract} (Figure \ref{fig:booking}): A user can book a trip (a hotel and a train) by calling the \texttt{order()} function of this contract.
\end{itemize} 

The Booking smart contract calls the other two contracts.  Intuitively, it is just like calling another computer program inside a computer program. However, the advantage of running this application on the blockchain is that in the case one of the two bookings fails even though the other was \texttt{order()} function call will fail and, as a result, reverting the successful booking. The user will not lose any money. In the traditional deployment of this application on a non-blockchain environment, it would be more complex to the revert the user's successful booking.

The use of a Turing-complete language like Solidity for  smart contract networks and the capability for contracts to call one another open limitless creativity when it comes to application. For any computer application in the real world, in theory, we can develop an equivalent version to run on the blockchain. This is why, with the birth of Ethereum and recent smart contract network alternatives, we have been witnessing many businesses enter the blockchain space, most notably in the finance field with Decentralized Finance (DeFi).

\section{Blockchain Scalability}
Every transaction is broadcast to the whole network. So is every block. Block validation takes time and efforts too. Due to the chain topology of the blockchain, the fact that only one block can be the next node of the chain leads to many completing blocks being wasted, effectively reducing transaction throughput. On the other hand, we want blockchain to be the universal computer for every one, for every application, if at all possible. Unfortunately, scalability remains a top challenge of blockchain technology \cite{9133427}.

\subsection{The Blockchain Trilemma}
Blockchain is aimed at three goals: decentralized, secure, and scalable.  They cannot be all perfectly realized, at least  according to Ethereum's founder, Vitalik Buterin, who originated the term ``Scalability Trilemma'' for blockchain \cite{scalabilitytrilemma}: 
\begin{itemize}
\item \underline{Decentralized}: Set to provide trustless computing, Blockchain does not rely on a central point of control. It needs to be decentralized such that nodes participate autonomously and equally to each other.
\item \underline{Secure}: The blockchain must operate as expected, robust to malfunctions and attacks. As we discussed in the previous section, Blockchain is a Byzantine Fault Tolerant system. The more  failures of a large threshold of nodes it can sustain, the better security it provides.
\item \underline{Scalable}: Meant to be a ``world'' computer for all people to run all applications, Blockchain should scale with increasingly growing amounts of transactions. This is in terms of both storage and computation demands.
\end{itemize}
The decentralization goal requires  as many nodes as possible to participate in the block validation. Having more validator nodes, however, leads to more difficulty in maintaining consensus, thus security. Both decentralization and security goals  are achievable only with small-scale blockchain networks (network size or transaction volume). As such, the scalability goal is not met.

Blockchains are often forced to make trade-offs in this trilemma. Bitcoin offers excellent decentralization and security, but unattractive scalability. Due to the 10-minute block creation rule to guarantee security, the transaction processing is slow, on the order of 2-5 transactions per second (tps).  This is not practical for real-world payment at merchants. Traditional credit card systems such as Visa and Mastercard are 3 to 4 orders of magnitude faster. 

On the other hand, Solana, a smart contract network created in 2017 by Anatoly Yakovenko et al. \cite{solana}, sacrifices decentralization for scalability.  It is very fast. Solana does not use PoW consensus which is of course slow. It is based on a unique Proof-of-History (PoH) consensus algorithm, a variation of Proof-of-Stake (PoS). In theory, Solana's claimed throughput can be as high as 710,000 tps; the practically observed number is  about 50,000 tps, still much faster than Bitcoin. However, as pointed out by many, Solana is vulnerable to centralization. In Solana, like other PoS networks, the decision to add a block to the blockchain is made among a small subset of ``validator'' nodes. Since fewer nodes involve in the consensus decision, it is faster than Bitcoin where all nodes can be miners. The problem with Solana is that the Solana Foundation is the only entity developing core nodes (validators) on the blockchain. This means Solana has a central point of control that reduces the network’s overall decentralization. In comparison, several core node developers are building on Ethereum (e.g. Go Ethereum, OpenEthereum, Nevermind, and Besu). As of April 1, 2022, the number of Solana validator nodes is estimated to be around 1,100 nodes; in comparison, the PoS Ethereum network already has more than 200,000 nodes.

Comparing Ethereum (the PoS version) and Bitcoin, both offer excellent decentralization. Ethereum is faster, but the tradeoff is in security where Bitcoin is the superior, mainly due to PoW which  has a higher entry barrier for block generation and higher cost to attack. An attacker would need to acquire 51\%+ of the computational power in the network, whereas a PoS attacker would need to acquire 51\%+ of the money within that system. To get the computational power in PoW, not only the attacker needs money but also physical efforts to acquire hardware. This  external and physical factor  makes PoW less vulnerable to attacks.

\subsection{Layer-2 Scalability}
Changing the core design, whether the consensus mechanism, block structure (chain or DAG), or cryptographic methods, at layer-1 of a blockchain network has tradeoffs due to the scalability trilemma. In 2016, at the peak then of high Ethereum gas fee,  Joseph Poon and Vitalik Buterin introduced the approach of layer-2 scalability that applies to Ethereum, and, in theory, any layer-1 blockchain. The proposed solution, called Plasma \cite{plasma}, builds a high-throughput blockchain network anchored atop  the layer-1 blockchain  as follows: 1)  \underline{layer-2 transaction processing}: users transact on the layer-2 blockchain, hence very fast; and 2) \underline{layer-1 transaction finality}:  state information records of completed transactions are saved in the layer-1 blockchain, hence assuring security against dishonest transactions. 

For example, Polygon\footnote{https://polygon.technology} is  a layer-2 smart contract blockchain on top of Ethereum network as layer 1. Polygon started with the Plasma approach in the early stage and now is one of the most successful blockchain networks. It is noted that the idea of layer-2 scalability  was actually  applied in  the Lightning Network \cite{lightningnetwork}, created by Joseph Poon and Thaddeus Dryja in 2015. The scaling method used is called \emph{State Channels}. This, however, is suitable only to a payment network like Bitcoin (as a fast payment protocol on layer 2), but not to a general smart contract network. Another scaling method often referenced is  \emph{Sidechain} \cite{sidechain}, which is a much simpler and less secure version of Plasma. 

The invention of Plasma opened a new direction in blockchain scalability, leading to more advanced solutions such as Optimistic Rollups\footnote{https://docs.ethhub.io/ethereum-roadmap/layer-2-scaling/optimistic\_rollups/} and ZK Rollups\footnote{https://docs.ethhub.io/ethereum-roadmap/layer-2-scaling/zk-rollups/}, which are trending today \cite{Sguanci2021Layer2B}.  To explain the layer-2 scalability's concept and feasibility, let us describe how Plasma works below. We hope that this will be helpful to the reader in understanding recently emerging scalability methods.

\subsubsection{Plasma Scaling}
There are several variants of Plasma. For example, Plasma Cash \cite{https://doi.org/10.48550/arxiv.1911.12095} is a Plasma solution for non-fungible tokens (NFT).  We present a basic version of Plasma - the original proposal \cite{plasma}, which is for fungible assets below.

\underline{The Plasma Chain}: First, we need to build a separate blockchain network to serve as the layer-2. We refer to this as    Plasma Chain and to the layer-1 chain as   Root Chain. Any blockchain design can work for   Plasma Chain as long as it is fast and scalable; for example, Proof-of-Stake or Proof-of-Authority is a better choice than Proof-of-Work for the consensus mechanism. In the initial Plasma proposal,   Plasma Chain adopts the UTXO state model (Bitcoin-like). Although this model is not suitable for enabling smart contracts at layer 2, for simplicity we assume this model to  explain  the core idea of layer-2 scalability.

 Plasma Chain processes transactions and creates blocks as usual functionalities of the chain. 
However, there is an additional step for the validator nodes (those that can produce blocks on   Plasma Chain) after they have added each block to   Plasma Chain: need to save a record of it on   Root Chain. This is called an on-chain  ``block commit''  or ``checkpoint''. By ``on-chain" we mean layer-1 activity, whereas ``off-chain" we mean layer-2. Adding a block to Plasma Chain provides its finality on Plasma Chain. Committing this block to Root Chain provides its finality on Root Chain, which is the ``finalized'' finality. The latest checkpoint is the proof  that all transactions  (and the funds) are permanent  up to this point. Blocks are committed on-chain by interacting with a smart contract on  Root Chain. There is also an entity, called Plasma Operator, which is watching events from this smart contract and will respond accordingly on  Plasma Chain. 

\underline{The Root Contract}: 
We need to create a smart contract on  Root Chain; let us name it  Root Contract. It provides the following functionalities:
\begin{itemize}
\item \texttt{Block Submission}:  Root Contract maintains a list of Plasma block headers, each  essentially consisting of the Merkle root of the corresponding original block and the time it is added to the list;  transactions are not included. The contract has a public function for inserting such a block into this list. This function is called by Plasma Chain's validator nodes after they have validated a block; alternatively, this can be called by Plasma Operator who watches block insertions on Plasma Chain. It is noted that because   Root Contract simply saves the headers of the Plasma blocks, not the actual transactions, it cannot know by itself their validity (honest or malicious purpose). 
\item \texttt{Fund Deposit}: Bob needs  some fund in his account before doing any transfer on   Plasma Chain. The contract has a public function for anyone like him to deposit this fund. Once this fund is deposited on   Root Chain, an event will be emitted to notify   Plasma Operator who will mint a new UTXO with the corresponding amount of fund on   Plasma Chain for Bob. The amount of fund in circulation on  Plasma Chain is the total amount of all deposits (minus withdrawn funds if any).
\item \texttt{Fund Withdrawal}: Alice  can withdraw fund from  Root Chain. The contract has a public function to allow so, which asks her to provide the proof for the fund used to withdraw.  A withdrawal must correspond to some unspent UTXO on Plasma Chain. The fund proof  includes the position of an unspent UTXO belonging to Alice on   Plasma Chain and the Merkle proof for this UTXO in its corresponding Plasma block. Because the contract cannot tell instantly  whether this UTXO is indeed unspent on   Plasma Chain, the  withdrawal is not immediate. It has to wait  a dispute  period, e.g., 7 days, during which anyone can challenge. If the challenge is valid, the contract will revert the withdrawal. 
\item \texttt{Fraud Proofs}:  The contract has a public function to allow anyone to challenge the validity of a malicious block committed from   Plasma Chain or a withdrawal request within its dispute period. In the case of challenging Alice's withdrawal request above, if Bob observes that the UTXO used in the withdrawal is also spent on the Plasma Chain, he will provide the position of this invalid UTXO on  Plasma Chain and the Merkle proof of its existence there  as input to the withdrawal-challenge function. The contract will see if this proof matches the corresponding block record in the Plasma block list of the contract. If matching, Alice's withdrawal will be reverted.
\end{itemize}

\underline{Example}: Consider a Plasma Chain on top of Ethereum for people to make ETH payments. 
\begin{enumerate}
\item Alice deposits 10 ETH to   Root Contract on   Root Chain. As a result,   Plasma Operator will mint 10 ETH for her on   Plasma Chain (this ETH on   Plasma Chain is actually a wrapped version of the Ethereum ETH). At this time, the Plasma blockchain consists of only 1 UTXO: 
\[
UTXO~1: \emptyset \rightarrow Alice: 10
\]
\item On  Plasma Chain, Alice transfers 5 ETH to Bob. The new blockchain state is:
\begin{align*}
spent: UTXO~1: \emptyset \rightarrow Alice: 10\\
UTXO~2: Alice \rightarrow Alice: 5\\
UTXO~3: Alice \rightarrow Bob: 5
\end{align*}
\item Bob then transfers 3 ETH to Charlie. The new blockchain state is:
\begin{align*}
spent: UTXO~1: \emptyset \rightarrow Alice: 10\\
UTXO~2: Alice \rightarrow Alice: 5\\
spent: UTXO~3: Alice \rightarrow Bob: 5\\
UTXO~4: Bob \rightarrow Bob: 2\\
UTXO~5: Bob \rightarrow Charlie: 3
\end{align*}
\item Charlie transfers 2 ETH to Alice. The new blockchain state is:
\begin{align*}
spent: UTXO~1: \emptyset \rightarrow Alice: 10\\
UTXO~2: Alice \rightarrow Alice: 5\\
spent: UTXO~3: Alice \rightarrow Bob: 5\\
UTXO~4: Bob \rightarrow Bob: 2\\
spent: UTXO~5: Bob \rightarrow Charlie: 3\\
UTXO~6: Charlie \rightarrow Charlie: 1\\
UTXO~7: Charlie \rightarrow Alice: 2
\end{align*}
\item At this time, on   Plasma Chain, Alice has 7 ETH (from UTXO 2 and UTXO 7), Bob has 2 ETH (from UTXO 4), and Charlie has 1 ETH (from UTXO 6). Note that the above transactions were included in Plasma blocks of  Plasma Chain and their headers have been saved in  Root Contract.   
\item Bob  requests to withdraw  2 ETH (calling the withdrawal function of   Root Contract on   Root Chain). He inputs to this function UTXO 4 as the source for the fund. The withdrawal request is pending for 7 days. During these 7 days, no one challenges this request because UTXO 4 is not spent on   Plasma Chain during the dispute period. Therefore,   Root Contract sends 2 ETH (of   Root Chain) to Bob. It is noted that Bob did not have to deposit fund on Ethereum in order to withdraw.
\item Alice requests to withdraw  5 ETH using UTXO 3. During the 7-day dispute period,  Charlie who watches Plasma Chain  observes that  UTXO 3  was spent on  Plasma Chain. He will challenge the withdrawal by submitting the Merkle proof of this UTXO 3 to Root Contract. This proof is valid, thus canceling Alice's withdrawal. 
\end{enumerate}

It is important  that those users who have fund on   Plasma Chain should  watch the chain frequently to make sure their funds are safe. This requires downloading the chain and verify its correctness. If a user detects or suspects something wrong, the user’s wallet (software) will automatically request to withdraw funds. 

To avoid spammers and those submitting irresponsible withdrawals while encouraging fraud reporting, one can design   Root Contract such that each withdrawal request must include a penalty bond that will be collected to reward the challenger in the case of bad withdrawal. To enable fast withdrawals (7 days is too long), a Plasma solution can involve Liquidity Providers who are incentivized to  advance the fund to the withdrawers while taking the risk of bad withdrawals. A solution, e.g,. Polygon, can also require that Alice burn the fund on Plasma Chain before requesting to withdraw it on Root Chain; she needs to submit the proof of this burn. 

The on-chain block commit in Plasma is the key difference between it and the Sidechain scaling approach \cite{sidechain}; the latter is often mistakenly considered the same as Plasma but it is very different. Sidechain also has a smart contract like  Root Contract with functions for deposits and withdrawals, but does not have block commits. It is simpler but a  major cons is that the sidechain can stop producing blocks and lock everyone’s funds up forever. Sidechain is thus much less secure. With Plasma, the block list in the Root Contract is the proof that users have their funds and thus can withdraw them.

\subsubsection{Rollups Scaling}
The Plasma approach is more suitable for token transfer transactions, but not for smart contracts. The Rollups approach \cite{rollups} was born to be general-purpose. The layer-2 blockchain in Rollups can run smart contracts. For example, one can  run an EVM inside the layer-2 chain, allowing existing Ethereum applications to migrate to Rollups without re-writing the smart contract code.

Rollups can be considered a hybrid Plasma approach. Plasma keeps all the transaction data off-chain and, as such,   Root Chain cannot verify Plasma transactions, leaving room for  Plasma Chain to do things maliciously. In Rollups, part of transaction data is saved on the Root Chain in addition to block headers. As a result,   Root Chain can verify transactions too, thus providing an additional layer to enhance security and decentralization. It is noted that Rollups does not save \emph{all} transactions on   Root Chain because doing so makes the Rollups chain meaningless; it does not do any scaling. If Rollups saved none of transaction data, it would become Plasma. To reduce the amount of transaction data saved on   Root Chain, it saves only the information necessary to verify transactions. Transaction data involving state storage remains on the Rollups chain.

There are two main Rollups approaches: Optimistic Rollups and Zero-Knowledge (ZK) Rollups. The former resembles Plasma in that it also uses fraud proofs to challenge invalid fund withdrawals and invalid layer-2 transactions. ZK Rollups is more disruptive in that it allows instant withdrawals. 
\begin{itemize}
\item \underline{Optimistic Rollups}:
The name ``optimistic'' comes from the assumption in this approach that  the transaction data submitted to   Root Chain is correct. After the Rollups chain commits a batch of transactions to   Root Chain, they will be considered permanently finalized if no one submits a fraud proof to challenge any transaction. Whenever a fraud proof is submitted, the suspicious transaction will be re-validated: it will be replayed on  Root Chain using the block state and transaction data information already saved in Root Contract. The replay of such transaction is similar to that in the transaction verification procedure of Ethereum. Noticeable implementations of Optimistic Rollups include  Optimism\footnote{https://www.optimism.io} and Arbitrum\footnote{https://offchainlabs.com/}.
\item \underline{ZK Rollups}:
ZK Rollups  leverages a cryptographic  method called zk-SNARK (Zero-Knowledge Succinct Non-Interactive Argument of Knowledge) \cite{chen2022review}. A zk-SNARK is a cryptographic proof that allows one party to prove that it possesses certain information without revealing that information. The verification of the proof is quick and cheap. When a batch of transactions are to be committed on   Root Chain, a zk-SNARK proof is computed for this data to prove its validity and sent along to   Root Chain.   Root Contract  verifies this proof on   Root Chain when receiving a withdrawal request; if valid, the fund is released immediately.  Noticeable implementations of ZK Rollups include dYdX\footnote{https://dydx.exchange},  Loopring \cite{loopring}, zkSync \cite{zksync}, and ZKSpace\footnote{https://zks.org}. 
\end{itemize}
To understand ZK Proofs, suppose that Alice wants to prove to Bob her knowing of a value $x$  such that $f(x)=output$ for a given $output$. Can she do that without disclosing value $x$? For example, can Alice provide a proof that she knows a secret value having a given SHA256 hash without revealing this secret? This is called a zero-knowledge proof. A related example is the well-known Yao's Millionaires' Problem \cite{Lin2005}: can two millionaires, Alice and Bob,  know who is richer without revealing their actual wealth? Mathematically put, with two numbers $a$ and $b$, can we determine whether $a \le b$ without revealing the actual values of $a$ and $b$? 

zk-SNARK is a method for computing ZK proofs. First, assume that  we can write a computer program to implement a boolean function $C(output,x)$ that returns true if and only if $f(x)=output$. For example, if $f$ is SHA256:

\begin{verbatim}
boolean function C(output, x) {
   	return (SHA256(x) == output);
}
\end{verbatim}

A zk-SNARK is a set of three functions, $Generator()$, $Prover()$, and $Verifier()$, defined as follows. 
\begin{align}
Generator(\lambda, C) \rightarrow (pk, vk)\\
Prover(pk, output, x) \rightarrow prf\\
Verifier(vk, output, prf) \rightarrow \{true, false\}
\end{align}
\begin{itemize}
\item $Generator()$: this is called the key generator. It takes as input a secret parameter $\lambda$ and program $C$ and outputs a pair of keys called a ``proving key'' ${pk}$, and a ``verification key'' ${vk}$. These keys are publicly known. It is noted that the secret parameter $\lambda$  must be known to no one except the generator.
\item $Prover()$: this is called the prover: Alice calls this function  taking as input the proving key ${pk}$, the public value $output$, and  her secret value $x$  that she wants to prove that $f(x) = output$. This function will output a value called ``proof" $prf$. Alice will send this proof to Bob.
\item $Verifier()$: this is called the verifier: Bob uses this function to take as input the verification key $vk$, the public value $output$, and the proof $prf$ he received from Alice. This function returns true iff the proof is correct; i.e., the prover knows a value $x$ satisfying $f(x)=output$.
\end{itemize}

As another example, suppose that Alice wants to transfer tokens of some ERC-20  cryptocurrency to somebody. Using the standard ERC-20 smart contract, the public sees the account balance of Alice, ${balance}$, and the amount she sends, ${value}$. In many cases, it is desirable to hide these numbers. For this purpose, we can implement a smart contract that  makes public only the following hashes of these numbers, ${balanceOld}$ = ${SHA256(balance)}$, ${sentValue}$ = ${SHA256(value)}$, and ${balanceNew}$ = ${SHA256(balance-value)}$. Knowing these hashed values,  the miner (Bob) cannot know the raw values, ${balance}$ and ${value}$, but can  still verify whether the transfer is valid. In this example, Alice is the prover and miner Bob is the verifier. The corresponding program code to define the logic for this verification, which is input into zk-SNARK, is as follows:

\begin{verbatim}
boolean function C(output, x) {
   	return (x.balance >= x.value  
		      && SHA256(x.balance) == output.balanceOld
		      && SHA256(x.value) == output.sentValue
		      && SHA256(x.balance-x.value) == output.balanceNew);
}
\end{verbatim}
Here, ${output}$ is the object consisting of the three hashed values that Bob observes and ${x}$ is the secret information about the sender's balance and value sent. With this program code $C$, the key generator will take it as input, together with a random parameter $\lambda$, to generate a proving key $pk$ and a verification key $vk$. Alice and Bob use these two keys to prove and verify as above.

There is tradeoff between Optimistic versus ZK Rollups. Due to mathematical complexity, generic constructions for ZK protocols  are too expensive to be used in practice. Thus far, it has  been suitable for only  a few specific applications such as payments and token exchanges, like what is mainly served by Plasma. Optimistic Rollups, on the other hand, thanks to its simplicity, supports layer-2 smart contracts better. It, however, requires more storage in  Root Contract (data needed to replay transactions for verification purposes). In contrast, with ZK proofs that can readily verify transactions, ZK Rollups requires less storage for  Root Contract.

\section{Blockchain Interoperability}
Existing  blockchain networks are each on their own island isolated from one another. Bitcoin users can only transact with other Bitcoin owners, but not with Ethereum users. Decentralized applications on Ethereum cannot make calls to those on other blockchain networks. Data on one blockchain cannot be shared outside either. This  is analogous to the early days of Internet, where different ``internets'' (networks) were developed independently to serve their own purposes or groups of users. They adopted different technologies and architectures that do not speak the same language. However, the Internet today is universally interoperable in that even though it consists of many Internet providers' networks, any two computers or applications regardless of where they belong can communicate with each other. 

Interoperability between the chains must be a top priority for blockchain. This should be seamless so that one should focus on the logic of the application without having to worry about which underlying blockchain technology stacks to use. Imagine the complexities that would arise for a supply-chain company if it runs the product tracking application on a blockchain and the payment application on another blockchain, and these two blockchains are not compatible.

At the least, we should enable interoperability for digital assets. We should be able to transfer or exchange assets between different networks without  intermediaries such as a centralized crypto exchange. This would allow a Bitcoin user to pay Bitcoin to a merchant that runs its Point-of-Sale software built on Ethereum.  This would benefit immensely decentralized finance (DeFi) applications that would be able to tap into all populations of users who own various types of assets. The next level of interoperability is for cross-chain exchanges of arbitrary data. This would enable smart contracts and applications  on different blockchains to communicate and share information. This kind of interoperability is of course much more difficult to achieve.

Efforts to realize blockchain interoperability remain fragmented. Protocols, however, have taken shape into three main approaches: Atomic Swap, Chain Bridge, and Chain Hub.
\subsection{Atomic Swap}
Atomic Swap \cite{10.1145/3212734.3212736} is a simple solution for two users to swap assets without involving any third party. They can be on the same chain or different chains. Suppose that Alice wants to transfer some asset X to Bob   who in return transfers some asset Y to her. In a naive scenario, Alice will just send X to Bob and expects him to send Y to her. The problem is, in the real world, Bob could just take her asset and run away. 

Atomic Swap guarantees that the exchange succeeds or else, nothing happens without either side losing asset. It works as follows. Alice and Bob each need to create a Hash-Time Locked Contract (HTLC) \cite{lightningnetwork} to deposit their respective asset. Specifically, \underline{Alice will do}:
\begin{enumerate}
\item Generate a secret key $k_{Alice}$. Only she knows it at this time.
\item Compute a crypto-hash value of this key, $m = H(k_{Alice})$. The hash function $H$ is known to Bob. 
\item Create a Hash-Time Locked smart contract (HTLC) on her chain to deposit asset X with a lock and an expiration time. This HTLC has a function to unlock X if it is called before expiration  and input with a key $k$ such that $H(k) = m$. 
	\begin{itemize}
	\item If asset X is unlocked, it will be transferred to the caller.  
	\item If X remains locked at expiration time, it will be returned to Alice.
	\end{itemize}
\item Send the hash value $m$ to Bob.
\end{enumerate}

On his side, \underline{Bob will do}:
\begin{enumerate}
\item Create a Hash-Time Locked smart contract (HTLC) on his chain to deposit  asset Y with a lock and an expiration time. This HTLC has a function to unlock Y if it is called before expiration  and input with a key $k$ such that $H(k) = m$. This value $m$ is the hash value sent from Alice.
\item Wait until the above unlock function is called and succeeds.
	\begin{itemize}
	\item If asset Y remains locked at expiration time, it will be returned to Bob.
	\item Else, the input key $k$ must equal the secret key of Alice, $k_{Alice}$. Therefore, Bob knows this private key. He will call the HTLC of Alice inputting this key $k= k_{Alice}$ to unlock  asset $X$  and have it transferred to him.
	\end{itemize}
\end{enumerate}

Atomic Swap will not do anything if Alice does not claim asset Y on Bob's contract, because if so Bob has no knowledge of her secret key to claim asset X on Alice's contract. If Alice does claim, Bob will know this key and claim his part too. No third party is involved here. On the other hand, Atomic Swap is not instant. It depends on the actions of Alice and Bob. Alice must send the hash value $m$ to Bob for him to set up his smart contract. She must then by herself contact his smart contract and vice versa.

\subsection{Chain Bridge}
While Atomic Swap is for swapping assets, Chain Bridge enables transfers of assets cross chains. To  illustrate its idea and feasibility, suppose that we want to bridge a smart contract blockchain X (token USDX) with a smart contract blockchain Y (token USDY).  
A basic Chain Bridge solution needs to write two smart contracts, one on X and one on Y. The bridge  is owned by an entity called Bridge Operator, who  watches events emitted from these contracts. Bridge Operator also has a liquidity pool $LP_X$ of $n_X$ USDX on X and a liquidity pool $LP_Y$ of $n_Y$ UDXY on Y. 

Suppose that Alice on chain X wants to transfer 10 USDX to Bob on chain Y. For simplicity, 1 USDX = 1 USDY and so he will receive 10 UDXY. The transfer happens as follows:
\begin{itemize}
\item \underline{On Chain X}: Alice calls the contract on X to deposit 10 USDX to the liquidity pool $LP_X$ on X. The new pool amount will become $n_X:= n_X + 10$.
\item \underline{Bridge Operator} detects this deposit and does the step below.
\item \underline{On Chain Y}: Bridge Operator calls the contract Y to transfer to Bob 10 USDY from the liquidity pool $LP_Y$. The new pool amount will be $n_Y := n_Y - 10$.
\end{itemize}

Since X and Y are existing chains in which Bridge Operator has no authority to mint assets, the liquidity pools are needed to provide instant liquidity for the transfer. The reserve amounts $n_X$ and $n_Y$ set the maximum amount one can transfer to X and Y, respectively. Thus, the more reserves, the more transfer volume is allowed. One can be creative by encouraging Liquidity Providers to contribute to these pools. 

In the case that Bridge Operator owns one of the two chains, say chain X, we do not need liquidity pool $LP_X$. In place of $LP_X$, Bridge Operator simply mints new USDX to the receiver anytime receiving a transfer from chain Y. Similarly, Bridge Operator burns USDX of the sender when needing to transfer it to chain Y. This is the solution often used when designing a new blockchain network that wants to bridge with an existing blockchain (e.g. Ethereum, so that the new network can host a wrapped version of ETH).

A challenge with  Chain Bridge is how to ensure security given the role of Bridge Operator \cite{https://doi.org/10.48550/arxiv.2101.06000}. For maximal security, Bridge Operator needs to be decentralized; ideally, it can itself be a blockchain network. However, that would lead to implementation complexities. In fact, no bridging solution has adopted such a method fully. One can resort to the cryptographic method of secure multi-party communication to partially decentralize the role of Bridge Operator, as in the Multichain framework\footnote{https://multichain.org/}, but to date  weak security remains the biggest concern for Chain Bridge. Many hacks targeted bridge solutions, most noticeable being the attack on Axie Infinity just this year (March 2022) incurring a loss of 600+ million USD. 

\subsection{Chain Hub}
Bridging is the interoperability solution to make two blockchains talk  to each other. If there are $n$ blockchains, we would need $n(n-1)/2$ bridges to enable any two chains to communicate directly. Chain Hub is an approach that builds the Internet of blockchains by providing a ``hub'' connecting to all the blockchains and dedicated to passing messages between them. More than that, this hub itself is a blockchain network, thus providing maximal decentralization and security.  Cosmos \cite{cosmos}, Polkadot \cite{DBLP:journals/corr/abs-2005-13456}, and Avalanche \cite{avalanche} are major solutions adopting this  approach. They call the ``hub'' by different names (Relay in Polkadot and Avalanche, or Hub in Cosmos).

\begin{figure}[t]
\centering
\includegraphics[width=\textwidth]{ 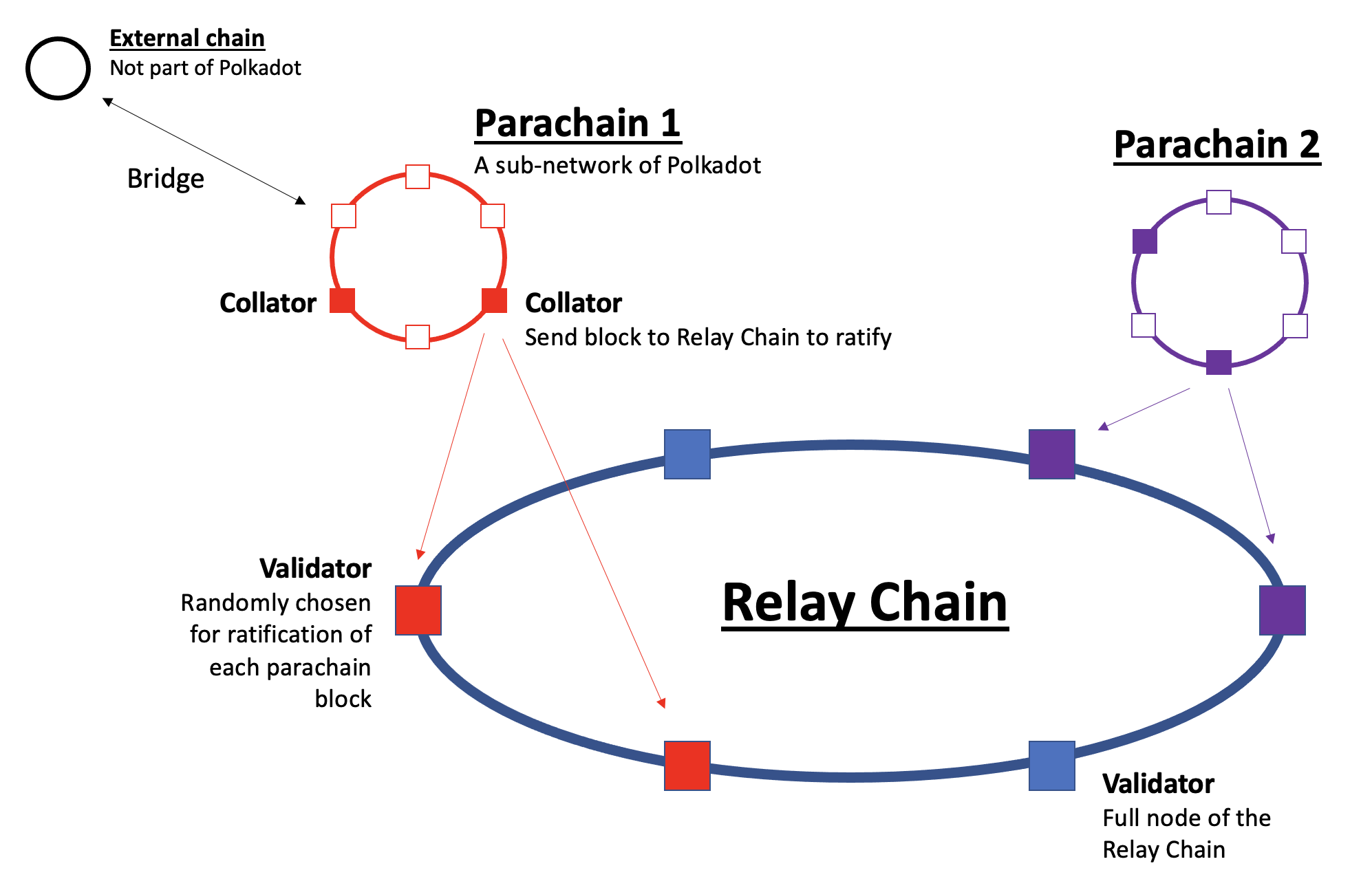}
\caption{Polkadot network: Blockchains (parachains) communicate with each other via Relay Chain. Parachain consensus is ensured by the Collators who are the validators of the parachain. Inter-parachain consensus is ensured by the Validators who are nodes on Relay Chain.}
\label{fig:polkadot}
\end{figure}

For example, consider Polkadot \cite{polkadotwhitepaper}, whose architecture is illustrated in Figure \ref{fig:polkadot}. Polkadot is a network of heterogeneous blockchain shards called ``parachains''.  These chains connect to and are secured by a chain called Relay Chain; this is the hub of Polkadot. Existing blockchains or those not of Polkadot network are called external networks which can talk to any parachain via bridges. 
There are four main roles for Polkadot keepers: validators, nominators, collators, and fishermen.

\begin{itemize}
\item \underline{Validators}: They must be among the nodes that form   Relay Chain. Once new blocks have been validated in their parachains,  they must be ratified on   Relay Chain. First, a subgroup of validators is chosen randomly to ratify each new parachain block. This results in a new block to add to Relay Chain. This block will be validated on Relay Chain as usual by all the validators.
\item \underline{Nominators}: They are stake-holding parties who risk capital to nominate nodes to become validators. Nominators get earnings if their nominees are chosen as validators. The method to choose validators from nominations is based on Nominated Proof of Stake (NPoS) consensus \cite{DBLP:journals/corr/abs-2005-13456}. In some sense, the validators are similar to the mining pools of current PoW blockchains and the nominators  are similar to the miners who join these pools.
\item \underline{Collators}: They must be among the parachain nodes. On their parachain, they author new blocks and execute transactions as usual (like miners in PoW blockchains or validators in PoS blockchains). In addition, as collators, they provide validators with valid parachain blocks (and zero-knowledge proofs) as candidate blocks to ratify on Relay Chain. We can think of collators as ``local helpers'' of validators on each parachain.
\item \underline{Fishermen}: They are ``bounty hunters'' who monitor Relay Chain and parachains to report irregularities committed by the nodes. They are rewarded by submitting a timely proof showinng that at least one bonded party misbehaved. The fishermen are an additional layer for enhancing the network security.
\end{itemize}

Polkadot can connect a set of independent blockchains while providing pooled security and trust-free cross-chain transactability, which is thanks to  Relay Chain with contributions from the above players. However, a Chain Hub solution like Polkadot requires building  blockchains  from scratch, which must use the same development framework (e.g., Substrate\footnote{https://www.parity.io/technologies/substrate/}  in Polkadot or Tendermint \cite{tendermint} in Cosmos) and abide a shared communication protocol. As such, a blockchain network adopting Chain Hub cannot interface with existing blockchains or those using non-compatible designs. Chain Hub is therefore called a layer-0 blockchain interoperability solution. In the future, one  hopes that Chain Hub will be successful and widely adopted. When that happens, we will realize the true vision of Blockchain being a universal computer or the next-generation Internet.

\section{Conclusions}
This chapter has presented  how blockchain works fundamentally, together with selective case studies, methods, and challenges, that help the reader understand this technology quickly to be sufficiently ready for further adventures. The coverage includes what blockchain is, its architecture and components, how it works for Bitcoin with Proof-of-Work consensus, the view of smart contract blockchains as universal computers, and open challenges in scalability and interoperability, the top-2  priorities for blockchain technology. 
It should become now clear that there is no limit in potential applications of blockchain  and emerging business models that otherwise are not feasible with conventional non-blockchain computing. However, despite its promise, blockchain technology is still in its infancy. Like the evolution of the Internet, it takes time for a new technology to mature and be widely accepted by traditional businesses. Technically, besides the foremost importance of scalability and interoperability, many other challenges remain to address as we go more deeply into each component of the blockchain architecture: how to optimize the peer-to-peer networking layer, innovate consensus mechanisms to be eco-friendly, incentivize and evaluate contributions to the security and decentralization of the blockchain,  develop smart contracts that are bug-free, enable decentralized finance for everybody, and apply effectively to other meaningful real-world  problems. All that makes the research and development of blockchain technology interesting.

\begin{acknowledgement}
Duc A. Tran's work for this chapter was partially funded by Vingroup Joint Stock Company and supported by Vingroup Innovation Foundation (VINIF) under project code VINIF.2021.DA00128. Bhaskar Krishnamachari's work was supported in part by the USC Viterbi Center for Cyberphysical Systems and the Internet of Things.
\end{acknowledgement}

%


\end{document}